\documentclass[reprint,amsmath,amssymb,aps,pre]{revtex4-1}

\usepackage[normalem]{ulem}
\usepackage{microtype}
\usepackage{hyperref}
\usepackage{bm}
\usepackage{xcolor}

\newcommand{\ud}{\mathrm{d}}

\usepackage{booktabs}

\usepackage{graphicx}
\usepackage{amssymb,amsmath,tabularx}
\usepackage{bm}
\usepackage[caption=false]{subfig}

\usepackage[normalem]{ulem}

\newcommand\SupplementaryMaterials{%
  \newcounter{sfigure}
  \renewcommand{\thefigure}{S\arabic{sfigure}}
  \newcounter{stable}
  \renewcommand{\thetable}{S\arabic{stable}}
  \newcounter{sequation}
  \renewcommand{\theequation}{S\arabic{sequation}}
  \stepcounter{sfigure}
  \stepcounter{stable}
  \stepcounter{sequation}
}

\begin{document}
\title{
Modeling COVID-19 dynamics in Illinois under non-pharmaceutical interventions
}

\author{George N.~Wong$^{1,\dagger}$, Zachary J.~Weiner$^{1,\dagger}$, Alexei V. Tkachenko$^{2}$, Ahmed Elbanna$^{3}$, Sergei Maslov$^{1, 4,5\ast}$, and Nigel Goldenfeld$^{1,5\ast}$}

\affiliation{$^1$Department of Physics, University of Illinois at Urbana-Champaign, Urbana, IL 61801, USA\\
$^2$Center for Functional Nanomaterials, Brookhaven National Laboratory, Upton, NY 11973, USA\\
$^3$Department of Civil Engineering, University of Illinois at Urbana-Champaign, Urbana, IL 61801, USA\\
$^4$Department of Bioengineering, University of Illinois at Urbana-Champaign, Urbana, IL 61801, USA\\
$^5$Carl R. Woese Institute for Genomic Biology, University of Illinois at Urbana-Champaign, Urbana, IL 61801, USA}
\author{\footnotesize{
$^\ast$ Correspondence to: \url{maslov@illinois.edu} and
\url{nigel@illinois.edu.}\\
$^\dagger$ These authors contributed equally to this work.}
}
\medskip
\date{\today}

\begin{abstract}
We present modeling of the COVID-19 epidemic in Illinois, USA,
capturing the implementation of a Stay-at-Home
order and scenarios for its eventual release.
We use a non-Markovian age-of-infection model that is capable of handling long and variable time delays without changing its model topology.
Bayesian estimation of model parameters is carried out using Markov Chain Monte Carlo (MCMC) methods.
This framework allows us to treat all available input
information, including both the previously published parameters of the epidemic and available local data, in a uniform manner.
To accurately model deaths as well as demand on the healthcare system, we calibrate our predictions
to total and in-hospital deaths as well as hospital and ICU bed occupancy by COVID-19 patients.
We apply this model not only to the state
as a whole but also its sub-regions in order
to account for the wide disparities in population size and density.
Without prior information on non-pharmaceutical interventions (NPIs), the model independently
reproduces a mitigation trend closely matching mobility data reported by Google and Unacast.
Forward predictions of the model provide
robust estimates of the peak position and severity and also enable
forecasting the regional-dependent results of releasing Stay-at-Home orders.
The resulting highly constrained narrative of the epidemic is able to
provide estimates of its unseen progression and inform
scenarios for sustainable monitoring and control of the epidemic.

\end{abstract}

\maketitle

On January 24, 2020, the second known COVID-19 case to be diagnosed in
the USA was reported in Chicago, Illinois.  Community transmission of
the disease was confirmed on March 8, 2020.
During the subsequent ten
days, the epidemic grew with a case doubling time of approximately 2.3
days, while
testing capacity was essentially fixed.
On March 21, 2020, a Stay-at-Home order was issued for the entire state of
Illinois and subsequently extended on March 31, 2020 and again on April 23,
2020. The order was lifted on May 30, 2020~\cite{Illinoisresponsewiki}.
Responsible relaxation of the mitigation of COVID-19 must
be informed by realistic and well-calibrated epidemiological modeling
of the outcomes of any scenarios under consideration---not just of the
resulting (increased) death toll but also of the stress placed upon the
healthcare system.
The purpose of this report is to present such an analysis.

A variety of modeling approaches are used by hospitals, public
health officials, and state governments. These range between
phenomenological models that use a curve-fitting procedure to match data, such as the daily death rate, and mechanistic
methods that model the trajectory of the epidemic as individuals transition
through several disease and healthcare-bound stages~\cite{keeling2011modeling,Murray2020,MA2020129,wu2020nowcasting}. Only
mechanistic models are able to make justifiable predictions
while accounting for changes in the epidemic environment,
such as the
imposition or relaxation of community mitigation efforts.  Of these,
compartmental models like the
Susceptible-Infectious-Recovered (SIR) models, and
Susceptible-Exposed-Infectious-Recovered (SEIR) extensions, are widely used.
Compartmental models describe how fractions of a homogeneous, well-mixed population progress through different states the disease, driven by interactions
between infectious and susceptible individuals.
In the simplest models, the dynamics is deterministic and the rates are constant in
time, but many variants and extensions exist and are widely used.

In order to be practically useful, models must be
calibrated to observed data~\cite{Fraser2007,chowell2017fitting,MA2020129,Wu2020}.
We calibrate the important dynamics of the model
to several simultaneous streams of empirical data
including total and in-hospital deaths, as well as hospital and ICU bed occupancy by COVID-19 patients.
To avoid biases resulting from non-uniform and non-constant testing
rates, which may be difficult to parameterize, we do not consider positive case data.
The resulting model is a description of the epidemic as it
progresses through the hospital system in Illinois;
as it is clear that a non-negligible number of COVID-19 deaths
occur outside the hospital environment (e.g., in homes and nursing
homes especially), we augment our model with an effective description of the net incidence
of deaths due to COVID-19.

There are many limitations to the types of models that we and others
use to describe COVID-19, and these have been explored extensively in
the literature, especially with regard to spatial structure~\cite{viboud2006synchrony}, superspreader events and individuals~\cite{LloydSmith2005,small2006super,bansal2007individual,kim2018agent},
and the structure of contact networks~\cite{dezsHo2002halting,rock2014dynamics,pastor2015epidemic}.
A geographical region as large as the state of Illinois is not
well-described as homogeneous, due to large variations in population
density and prevalence of infection between the Chicago metropolitan area and the more rural
regions in Central and Southern Illinois.
Indeed, most of the known cases and deaths to
date have occurred in the Chicago area (by roughly by an order of magnitude), so
in practice this region dominates our results for the state as a whole.

One cannot simply scale the results of models for the Chicago
metropolitan area to the entire state, however, because the
transmission characteristics depend on the frequency and duration of
contacts, which likely vary significantly among geographical regions.
For this reason, we simulate the dense urban areas and the three
sparser, more rural areas separately. We note that a more refined
treatment would account for transfers between these separate
populations, as well as transfers into and out of the state as a whole;
however, we do not currently model these processes. The number of cases in
individual rural areas is sufficiently small, due to the early
Stay-at-Home order, that a long phase of exponential growth in these
regions was largely avoided. As a result, these populations are not
well-described by continuum, deterministic models. Nevertheless, by
aggregating these populations, the numbers are large enough that
deterministic, exponential growth, at least at the early stages, was visible.

In this work, we describe an estimate of the rise and
fall of the epidemic within Illinois, taking into account the
modulation of the transmission parameters due to social distancing.  In the
following sections, we first describe our extension of the SEIR model, which
takes into account the long and variable delay times
reported in the literature.  After describing the procedure by which we
calibrate the model to data, we argue for the robustness of short term model predictions.
Finally, we present and discuss our predictions of the epidemic trajectory
through the Summer of 2020, including the effects of the
release of the Stay-at-Home order at the end of May.

\section{Model description}

To facilitate a general treatment of delay times in the COVID-19
epidemic, we implement a non-Markovian model that derives from the
classic Kermack-McKendrick age-of-infection
model~\cite{kermack1927contribution}. Age-of-infection models are
similar to compartmental models, such as
Susceptible-Infectious-Recovered (SIR) or
Susceptible-Exposed-Infectious-Recovered (SEIR) models, but allow
transition delays between entering sequential states to be drawn from
arbitrary probability density functions (see, e.g. ~\cite{Wallinga2006,
MA2020129}). We use non-Markovian models because their delays can be
defined by an arbitrary number of timescales, in contrast to the single
exponential rate parameter of compartment based models. Non-Markovian
models are thus, in principle, better able to reproduce the observed
signal delays between different states, e.g., the flattening of the
curve of hospital admissions compared to in-hospital deaths.

The use of deterministic rather than stochastic epidemic descriptions
is generally justified when the modeled populations are large enough
that relative daily changes are small and the number of individuals is
large relative to one.  This means that a deterministic compartmental
model has a self-consistency check, because once the epidemic size is
of order unity, a stochastic model allows the epidemic to die out; in
contrast, in a deterministic model an epidemic will continue to evolve
even with a unrealistic, fractional number of infectious individuals.
We will see that for some regions of Illinois, our estimates are at the
limit of validity of deterministic models.

We adapt the conventional age-of-infection model to include a number of
delayed healthcare system observables of the epidemic, such as the
number of patients in hospitals, Intensive Care Units (ICUs), and
ultimately, the number of deaths.

\subsection{Time-modulated age-of-infection model}\label{subsec:model-time-description}

The core of our model is the daily incidence (i.e., the number of newly-infected
individuals) in demographic (age) group $i$, $j_i(t)$.
This value determines the dynamics of susceptible individuals in that group $S_i(t)$ according to
\begin{align} \label{eqn:S}
    \frac{\ud S_i(t)}{\ud t}
    &= - j_i(t).
\end{align}
The incidence itself follows the renewal equation,
\begin{align} \label{eqn:delta-I-integral}
    j_i(t)
    &=  R_t \frac{S_i(t)}{S(t)} \sum_m \xi_{im} \int\limits_{0}^{\infty} \ud \tau \, K_\mathrm{serial}(\tau) j_m(t-\tau).
\end{align}
Here, $R_t$ is time-dependent effective reproduction number,
$S(t)=\sum_i S_i(t)$ is the total susceptible population,
$K_\mathrm{serial}(\tau)$ is the probability density function (PDF) of serial intervals,
and $\xi_{im}$ is the contact matrix. For simplicity, we assume $\xi_{im} = 1$, i.e.,
all demographic groups infect each other at the same rate.
We assume that $N_i$, the total number of individuals in the demographic group $i$, is
approximately constant for
the duration of the epidemic. In practice, this means our model neglects the effects of
births, deaths due to causes unrelated to COVID-19, and mobility of the population,
which is appropriate for the short time scales we model.
We further assume that individuals are only infected once, i.e., that the
duration of immunity to COVID-19 is longer than the timescale over which we simulate the epidemic.
Our simulations begin on a day $t_s$ (whose value we sample during parameter inference)
at which point we impose that ten people are spontaneously infected, setting $j_i(t)$ in proportion to the age
distribution of the population under study.

We parameterize the effective reproduction number $R_t$  in terms of the basic reproduction number $R_0$, a seasonal forcing factor $F(t)$, a mitigation
factor $M(t)$, and the susceptible population fraction $S(t)/N$ according to
\begin{align}
    R_t
    &= R_0 F(t) M(t) \frac{S(t)}{N} .
\label{eqn:Rt}
\end{align}
The formal dependence of $R_t$ on the total susceptible population $S(t)$
is corrected in Eq.~\ref{eqn:delta-I-integral} by a factor $S_i(t) / S(t)$
which accounts for possible variation of susceptibility between different demographic groups.
In our model, the homogeneous factor $M(t)$ accounts for all sources of mitigation,
including the effects of self-imposed isolation as well as government-mandated
Stay-at-Home (SAH) orders, school closures, etc.
We parameterize $M(t)$ as a piecewise cubic Hermite interpolating polynomial that
smoothly interpolates from 1 at $t_0$ to $M(t_1)$ at $t_1$ and is otherwise constant.
The mitigation factor $M(t_1)$, start time, and end time of the interpolation interval
are fitted by our algorithm.
We choose to parameterize $M(t)$ as a single transition (i.e., a single event) since this minimal model reduces the risk of overfitting to spurious trends. This choice is supported by the following observations:
the adoption of social distancing practices took place relatively rapidly over a one to two week period; and the Stay-at-Home order remained active after the initial transition, presumably suppressing the magnitude of changes to mitigation. We will observe that this choice is a sufficient approximation for the duration of the data we use for calibration.
We evaluate these claims more explicitly
Section~\ref{sec:evaluation-params}.

To model seasonal effects, we follow Ref.~\cite{Neher2020}, which estimates seasonal forcing from the observed
seasonal variability of positive tests in three other endemic coronaviruses. We thus adopt the
functional form
\begin{align}
    F(t)
    &= 1 + A_{\mathrm{SF}} \cos\left(  \frac{2 \pi(t - t_\mathrm{peak})}{365}
    \right),
\end{align}
where $A_{\mathrm{SF}}$ denotes the strength of the forcing and $t_{\mathrm{peak}}$ sets the day of the year when seasonal forcing is strongest. From Ref.~\cite{Neher2020}, we infer that seasonal forcing is strongest in the winter and set $t_{\mathrm{peak}} = $ January 16.

Ref.~\cite{Neher2020} finds evidence for $A_\mathrm{SF} = 0.2$; however, we account for uncertainty in this parameter
by sampling over $A_\mathrm{SF}$ during parameter inference.
Incorporating this uncertainty is critical: if mitigation
were only able to reduce the effective reproduction number to
roughly unity, then seasonal forcing
could drive a second wave of the epidemic.

For timescales of only a few months,
our parameterization of $R_t$ includes a degeneracy: a change in the parameter $A_{\mathrm{SF}}$ may be compensated by
adjusting the mitigation profile $M(t)$. The degeneracy is broken as data is collected over long timescales, since $M(t)$ models relatively instantaneous changes in infectiousness and $F(t)$ produces an explicitly year-long, periodic modulation.
Practically, this implies that $A_{\mathrm{SF}}$ may account for both
seasonal effects and concurrent slow variations in the mitigation factor.

\subsection{Model topology}

Due to limited and biased testing, neither the susceptible population
$S_i(t)$ nor the daily incidence of new infection cases $j_i(t)$ are directly
observable. Hence, we are forced to infer the dynamics of the epidemic
using lagging and indirect indicators.
These indicators include the total number of
hospitalized (but not critical) patients $H(t)$, the number of critically ill patients
currently in ICU beds $C(t)$, and $D(t)$, the cumulative number of
daily deaths in the hospital.

Our model topology assumes that all hospital deaths occur in ICU rooms.
In practice, this simplification would be invalid if either a significant number of individuals
die immediately upon entering the hospital (i.e., if the true delay distribution between
hospitalization and death is appreciably bimodal), or if the number of hospital decedents
exceeds the number of individuals who are admitted to the ICU.
In this sense, our inferred parameter values, e.g., the probability of a patient in critical care dying,
should not be interpreted as having real-world meaning, since the values accommodate approximations
in order to fit all input data.

Furthermore, by separating the observables from the incidence dynamics, our model supposes
that the hospitalization status of an individual does not affect their likelihood of infecting someone.
This choice is appropriate if the number of hospitalized individuals is a small fraction
of the total number of infected individuals, or if the delay between infection and hospitalization
is longer than the serial interval.

\begin{figure}[ht!]
\centering
\includegraphics[width=\columnwidth]{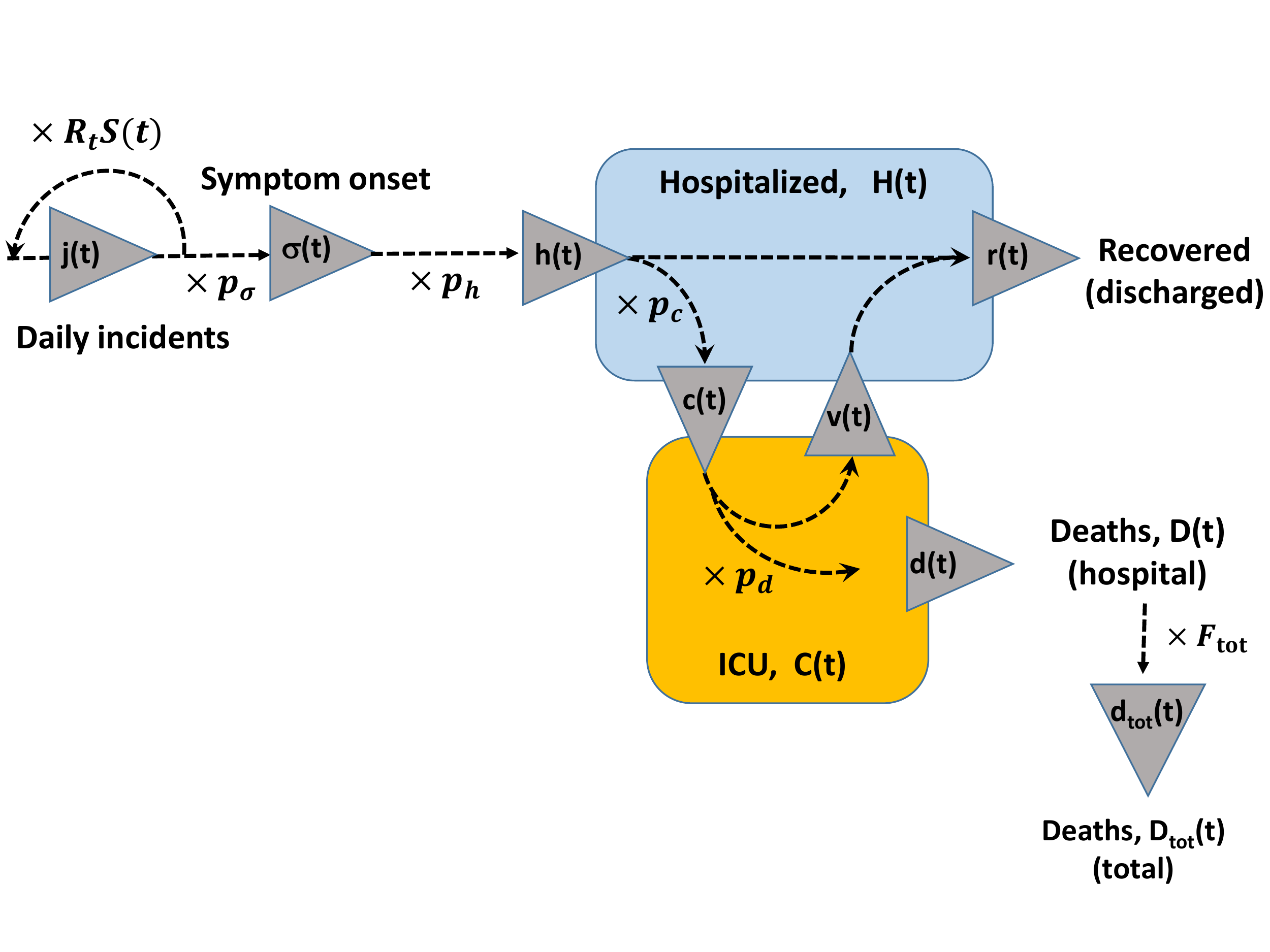}
\caption{
    The topology of our model along with the names of all flux and
    state variables: the daily incidence, $j_i(t)$; the daily number of
    newly symptomatic individuals, $\sigma_i(t)$; the number of daily
    admissions to all hospitals, $h_i(t)$; the daily number of patients
    discharged from all hospitals, $r_i(t)$; the daily number of patients
    transferred from the main floor of a hospital to its ICU, $c_i(t)$; the
    daily number of patients transferred from the ICU to the main floor of
    a hospital, $v_i(t)$; the daily number of deaths in hospitals, $d_i(t)$; and
    the daily number of deaths in and out of hospitals, $d_{\mathrm{tot}, i}(t)$.
    State variables are: the total number of occupied hospital beds (main
    floor) $H_i(t)$, and the total number of occupied ICU beds $C_i(t)$. The
    other parameters of the model are the fractions of infected individuals
    who ever become symptomatic, $p_{\sigma, i}$; the fraction of symptomatic
    individuals who are ever hospitalized, $p_{h, i}$; the fraction of
    hospital patients who ever get to ICU, $p_{c, i}$; and the fraction of ICU
    patients who will ultimately die $p_{d, i}$; and the multiplier, $F_{\mathrm{tot}}$
    that converts between hospital deaths and all deaths in the state, including those
    outside of the hospital system.
    For the sake of legibility, we suppress age-group indices in the diagram.
}\label{fig:model-topology}
\end{figure}

The dynamics of the system may be described by daily flux variables:
\begin{itemize}
    \item $\sigma_i(t)$, the number of infected individuals who become symptomatic
    \item $h_i(t)$, the number of daily admissions to all hospitals
    \item $r_i(t)$, the daily number of patients discharged from all hospitals
    \item $c_i(t)$, the daily number of patients transferred from the main floor of a hospital to its ICU
    \item $v_i(t)$, the daily number of patients transferred from the ICU to the main floor of a hospital, and
    \item $d_i(t)$, the daily number of deaths in ICU rooms.
\end{itemize}
We do not directly model deaths that happen outside of hospitals but instead
infer the ratio of these deaths to the hospital deaths during our fitting
procedure, as described below.
Figure~\ref{fig:model-topology} schematically depicts the topology of
our model along with the names of all flux and cumulative variables.

The dynamics of any flux variable $y(t)$ defined above may be obtained
from the variable $x(t)$ directly preceding it in the chain of events
shown in Fig.~\ref{fig:model-topology}:
\begin{align}
    y(t)
    &= p_{y}
        \int\limits_0^\infty \ud \tau \,
            K_{y}( \tau) x(t - \tau).
\end{align}
Here, $p_y$ is the proportion of individuals undergoing the transition $x \to y$ with time delays
distributed according to a probability density function
$K_y(t)$.
Note that Eq.~\ref{eqn:delta-I-integral} has the same structure except
that both the input $x$ and the output $y$ variables are given by the
daily incidence $j_i(t)$.

For the flux variables defined above, one obtains the following
equations. The number of infected individuals who become symptomatic is
\begin{align}
    \sigma_i(t)
    &= p_{\sigma, i}
        \int\limits_0^\infty \ud \tau \, K_\sigma(\tau) j_i(t - \tau),
\end{align}
where $p_{\sigma, i}$ is the (age-dependent) fraction of infected individuals who ever
develop symptoms and $K_\sigma(\tau)$ is the PDF of the incubation
period.
The fraction ${p}_{h, i}$ of symptomatic individuals who are ever admitted to the hospital is
\begin{align}
    h_i(t)
    &= p_{h, i}
        \int\limits_0^\infty \ud \tau \, K_h(\tau) \sigma_i(t - \tau).
\end{align}
The flux of hospital patients who become critically ill
and are admitted to the ICU is given by
\begin{align}
   c_i(t)
    &= p_{c, i}
        \int\limits_0^\infty \ud \tau \, K_c(\tau) h_i(t - \tau),
\end{align}
of which a fraction $p_{d, i}$ ultimately die according to
\begin{align}
    d_i(t)
    &= p_{d, i}
        \int\limits_0^\infty \ud \tau \, K_d(\tau) c_i(t - \tau).
\end{align}
The fraction $1 - p_{d, i}$ of ICU patients who stabilize and return to the general ward of the
hospital do so according to
\begin{align}
    v_i(t)
    &= (1 - p_{d, i})
        \int\limits_0^\infty \ud \tau \, K_v(\tau) c_i(t - \tau).
\end{align}
Both stabilized patients and hospital patients who recover
without requiring critical care are discharged, thus the influx of
recovered individuals due to hospital discharges is given by
\begin{align}
    r_i(t)
    &= \int\limits_0^\infty \ud \tau \, K_r(\tau) \left[
            (1 - p_{c, i}) h_i(t - \tau)
            + v_i(t - \tau)
        \right].
\end{align}

Finally, to approximately account for patients who die outside of the hospital system, we introduce $d_{\mathrm{tot}}$, a variable that tracks the total number of daily deaths both within and outside of the hospital. We connect total deaths to hospital deaths $d$ according to a prefactor $F_{\mathrm{tot}} \ge 1$ and delay time $\tau_{\mathrm{tot}}$ that reflects bureaucratic delays associated with issuing deaths certificate and publishing data on the Illinois Department of Public Health (IDPH) website. We observe that these bureaucratic delays manifest themselves in strong day-of-the-week effects.
\begin{align}
    d_{\mathrm{tot}, _i}(t)
    &= F_{\mathrm{tot}}
        \int\limits_0^\infty \ud \tau \, K_{\mathrm{tot}}(\tau) d_i(t - \tau).
\end{align}

In our simulations, we draw on clinical data to specify the age-dependent rates for
hospitalization, ICU admission, and death. We report the details of this severity model in Table~\ref{tab:severity-model}.
To account for differences between the literature values and what has occurred in Illinois, we
introduce age-independent prefactors for the
transition rates and fit them in our simulations.
Although the relative severity values we use may not be accurate,
in practice this choice does not affect our model dynamics.
Because the contact matrix $\xi_{im}$ is constant across all interactions, and because the susceptibility is not a function of demographic group, the demographic-aggregated statistics are insensitive to the relative demographic ratios.
Then, since we only fit data that has been summed over all age groups, we cannot observe any differences caused by inaccurate severity ratios.

The instantaneous occupation of hospital beds $H(t)$ and of ICU beds $C(t)$
may be obtained by integrating the incoming and outgoing fluxes as
\begin{align}
    H(t)
    &= \sum_i \int\limits_0^t \ud t' \left[
            h_i(t')-c_i(t')+v_i(t')-r_i(t')
        \right] \\
    C(t)
    &= \sum_i \int\limits_0^t \ud t' \left[
            c_i(t')-v_i(t')-d_i(t')
        \right],
\end{align}
while the cumulative numbers of hospital and total deaths are
\begin{align}
    D(t)
    &= \sum_i \int\limits_0^t \ud t' \, d_i(t') \\
    D_\mathrm{tot}(t)
    &= \sum_i \int\limits_0^t \ud t' \, d_{\mathrm{tot}, i}(t').
\end{align}

\section{Parameter inference}

We calibrate our model to data by sampling over the high-dimensional
model parameter space using a Markov chain Monte Carlo (MCMC) approach,
as has been done by many others for epidemics in general (see, e.g.
\cite{chowell2017fitting,smirnova2019forecasting}) and also for
COVID-19 (see, e.g. for applications in China~\cite{Ku2020}, Mexico
\cite{mena2020using} and Italy~\cite{gatto2020spread}). While standard
optimization techniques can also identify the best-fit model
parameters, we use MCMC because it produces an estimate of the global
posterior probability distribution. With the full distribution, we can
motivate bounds on parameter uncertainty, explore correlations between
parameters, and identify model idiosyncrasies. Access to the full
distribution also provides a direct means to marginalize over some
modeling uncertainties when forecasting the future trajectory of the
epidemic.

\subsection{Markov chain Monte Carlo methods}

Given a set of model parameters $\vec{\Theta}$ and data $\vec{\mathcal{D}}$,
the input to the MCMC sampler comprises a prior $\pi(\vec{\Theta})$ and a
likelihood function $L(\vec{\mathcal{D}} | \vec{\Theta})$. The sampler
computes the posterior probability $p(\vec{\Theta} |\vec{\mathcal{D}})$ for each point in
parameter space according to Bayes's theorem
\begin{align}
    p(\vec{\Theta} \vert \vec{\mathcal{D}})
    = \frac{L(\vec{\mathcal{D}} \vert \vec{\Theta}) \pi(\vec{\Theta})}{Z(\vec{\mathcal{D}})} .
\end{align}
The likelihood $L$ represents the
probability of observing the data $\mathcal{D}$ given a model with input parameters
$\vec{\Theta}$, and the prior $\pi$ encodes our expectation for the probability of a given
set of model parameters $\vec{\Theta}$.
Because we only compare points within the posterior distribution for
an individual set of data, we neglect the constant model evidence
by setting $Z(\vec{\mathcal{D}})=1$.
In practice, the MCMC sampling recovers the log of the posterior probability distribution
$H\equiv \ln p + \mathrm{const}$, which combines both type of inputs according to
\begin{align}\label{eqn:log-likelihood-2}
    H (\vec{\Theta} \vert \vec{\mathcal{D}}) = \ln \left[L(\vec{\mathcal{D}} \vert
    \vec{\Theta})\right]-\sum_\alpha  \frac{\left
    (\Theta_\alpha-\Theta^{(0)}_\alpha\right)^2}{2\Delta_\alpha^2}.
\end{align}
Here, the second term is the log of the prior over the model parameters;
for each parameter we either implement a Gaussian prior with mean expected value $\Theta^{(0)}_\alpha$ and variance $\Delta_{\alpha}^2$,
or we use a flat prior in which case the parameter does not appear in the sum.

This Bayesian framework enables a uniform treatment of all available
input information information, i.e., both observed time series data and
the parameters of the model. We determine the prior means by averaging
published clinical data weighted in proportion to the sample size of
each study. To account for differences between reported estimates of
parameters due to, e.g., possible variability of model parameters
between different locations, the tolerance parameters $\Delta_\alpha$
were estimated as \textit{unweighted} root-mean-square deviations of
the published data from their respective average values
$\Theta^{(0)}_\alpha$ across the different studies. As a result, our
procedure is flexible with respect to any local variability in model
parameters. If high quality local data on parameter values are
available for calibration, as might be the case for the duration of ICU
stays or severity models, these data can be used directly with small
values (or zero) for the respective tolerance parameters. By forcing
parameters with known values to be more tightly constrained, unknown
parameters will be automatically optimized with respect to all data
types, and we can thus increase the fidelity of our model calibration
result.

Many of our model outputs and data quantify daily incidences, e.g., the number of deaths per day. For these sorts of rate data, a Poisson likelihood estimator is appropriate.
For a
data point $d$ at time $t$, the Poisson likelihood is given by
\begin{align}\label{eqn:point-likelihood}
    L(\lambda|t, d) = \left(\frac{e^{-\lambda(t)} \lambda(t)^{d}}{d!}\right)^{1/T},
\end{align}
where the time-dependent rate $\lambda(t)$ is equal to the model output and $T$ is the correlation time for the data. The
likelihood over the full data set is the product over the
likelihoods for each individual data point; thus,
the total log-likelihood is given by
\begin{align}\label{eqn:data-log-likelihood}
    \ln \left[L(\vec{\mathcal{D}} |
\vec{\Theta})\right] = \sum\limits_i \dfrac{d_i \ln \lambda(t_i) - \lambda(t_i) - \ln \left(d_i!\right)}{T_i} .
\end{align}

In addition to being a well-motivated choice when comparing count data to rates,
the Poisson likelihood is appealing because its effective
uncertainty scales with the rate parameter.
Thus, unlike with the $L_2$ norm, a single parameter specifies
both the expectation value and the uncertainty of the measurement.
In practice, we found that evaluating likelihoods using the $L_2$ norm did not
significantly alter the qualitative features of our forecasts.

We divide by $T$ in Eq.~\ref{eqn:data-log-likelihood} because we
also calibrate against instantaneous hospital statistics, such as
occupancy in the general ward and in the ICU. These data sources have
natural correlation timescales: occupancies correspond to smoothed
averages since the majority of individuals who occupy a bed do so
continuously over several days. We set $T$ equal to the correlation
timescales $\Theta^{(0)}$ from our priors. In particular, we assume a
correlation of $6$ days for occupied hospital beds, and a correlation
of $12.75$ days for occupied ICU beds.
We set $T=1$ for the raw incidence data, i.e., for daily hospital deaths and for daily total deaths.

In Table~\ref{tab:sample-params}, we enumerate the model parameters we
sample over and list the bounds on those parameters' values. We also
describe the shape of any prior distributions we impose. In our model,
we use gamma distributions to describe delays, and specify the mean and
standard deviation for each distribution.  Here, the mean $\tau$ and
standard deviation $\sigma$ of a gamma distribution are related to the
standard shape and scale parameters by $k = \tau^2 / \sigma^2$ and
$\theta = \sigma^2/\tau$, respectively. We fix the serial interval
mean and standard deviation to $4$ and $3.25$ days
respectively~\cite{nishiura_serial,Du2020}, while our incubation time
distribution has fixed mean $5.5$ days and a standard deviation of $2$
days~\cite{incub1,incub2}. Parameters for all other delays are sampled.
\begin{table}[th]
\centering
\begin{ruledtabular}
\begin{tabular}{llll}
    {} &       bounds & $\Theta^{(0)}$ & $\Delta$ \\
    \midrule
    $R_0$                     &     $[1, 5]$ &             -- &       -- \\
    $t_s$                     &   $[35, 65]$ &             -- &       -- \\
    $t_0$                     &   $[60, 85]$ &             -- &       -- \\
    $t_1$                     &  $[70, 100]$ &             -- &       -- \\
    $M(t_1)$                  &  $[0.05, 1]$ &             -- &       -- \\
    $\mathrm{IFR}$            &   $[0.25\%, 1.8\%]$ &     $0.7\%$ &  $0.175^\star$ \\
    $\tau_h$                  &  $[0.5, 40]$ &          $6.5$ &      $2$ \\
    $\sigma_h$                &  $[0.5, 20]$ &            $4$ &      $2$ \\
    $\tau_{\mathrm{disch}}$   &  $[0.5, 40]$ &            $6$ &      $2$ \\
    $\sigma_{\mathrm{disch}}$ &  $[0.5, 20]$ &            $4$ &      $2$ \\
    $p_c$                     &  $[0.05, 4]$ &             -- &       -- \\
    $\tau_c$                  &  $[0.5, 10]$ &            $2$ &      $2$ \\
    $\sigma_c$                &  $[0.5, 10]$ &            $2$ &      $2$ \\
    $p_d$                     &  $[0.05, 3]$ &             -- &       -- \\
    $\tau_{d}$                &    $[4, 20]$ &           $12$ &      $3$ \\
    $\sigma_{d}$              &    $[1, 20]$ &          $8.5$ &      $3$ \\
    $\tau_\mathrm{rec}$       &    $[4, 20]$ &        $12.75$ &      $3$ \\
    $\sigma_\mathrm{rec}$     &    $[1, 20]$ &           $10$ &      $3$ \\
    $F_\mathrm{tot}$          &    $[1, 10]$ &             -- &       -- \\
    $\tau_\mathrm{tot}$       &    $[0, 10]$ &            $2$ &      $1$ \\
    $\sigma_{\mathrm{tot}}$   &    $[0, 10]$ &            $2$ &      $1$ \\
    $A_\mathrm{SF}$           &   $[0, 0.2]$ &             -- &       -- \\
\end{tabular}

\end{ruledtabular}
\caption{
    Specification of sampler parameters and their priors.
    The mean
    and standard deviation
    of a delay-time PDF $K_x$ are denoted
    by $\tau_x$ and $\sigma_x$ respectively, and probabilities $p_x$ denote
    the overall scaling of the age distribution $p_{x, i}$ as specified in
    Table~\ref{tab:severity-model}.
    In addition to the listed bounds on the various scaling factors $p_x$,
    we also enforce that the number of individuals drawn from a state does not exceed the
    total number presently in that state.
    We indicate strict bounds on parameter values
    and provide the mean $\Theta^{(0)}$ and standard deviation $\Delta$  for parameters' Gaussian priors where specified.
    Our priors on $K_h$, $K_r$, $K_c$, and $K_d$ are informed by Refs.~\cite{incub2,Verity2020,Bhatraju2020,Wuhan_clinical,icnarc},
    and use Ref.~\cite{systematic_review_ifr}
    to set our prior on $\ln \mathrm{IFR}$.\\
    {${}^\star$}We implement a Gaussian prior for $\ln \mathrm{IFR}$,
    with mean corresponding to $\mathrm{IFR} = 0.7\%$
    and variance of $\ln \mathrm{IFR}$ set to $0.175$.
    }
\label{tab:sample-params}
\end{table}

Finally, we implement MCMC sampling using the Python package
\textsf{emcee}~\cite{emcee}. To improve sampling efficiency, we use
ensemble move proposals based on the differential
evolution~\cite{terBraak2008}, differential evolution
snooker~\cite{Braak2006}, and kernel density~\cite{farr2014more}
proposal updates.

\subsection{Posterior distributions and data fitting}

We calibrate our model using data on hospital and ICU room
occupancy by COVID-19 patients, the number of daily deaths of COVID-19
confirmed patients in hospitals,
and the total number of daily deaths as publicly reported by the IDPH~\cite{IDPH}.
At the time of calibration, the
hospitalization and ICU data were not publicly available and were
provided to us by the IDPH.
The MCMC sampling procedure produces a high-dimensional posterior
probability. We use this posterior to identify the expectation values
and uncertainties for each parameter with respect to the model and the
data.

To summarize the posterior distribution in terms of epidemic
trajectories, we take a representative sample of parameters
$\vec{\Theta}_i$ according to their posterior probabilities. For each
$\vec{\Theta}_i$, we simulate the full course of the epidemic, and we
plot the resulting family of curves in aggregate. At every time point
we identify the median values of all measurable quantities (hospital
and ICU rooms, and deaths) as well as quantiles corresponding to 68.4\%
and 95.6\% confidence intervals. Because these quantiles are evaluated
independently at each point in time, the visually recognizable curves
do not correspond to actual epidemic trajectories.

\section{Modeling Results}
We used our age-of-infection model to describe the progression of
COVID-19 in Illinois during 2020.  We performed analyses for the state
and for four distinct localities of the state.  We also considered three
separate scenarios of social distancing in our modeling.

\subsection{Simulations for Illinois}

\begin{figure}[ht!]
\centering
\includegraphics[width=\columnwidth]{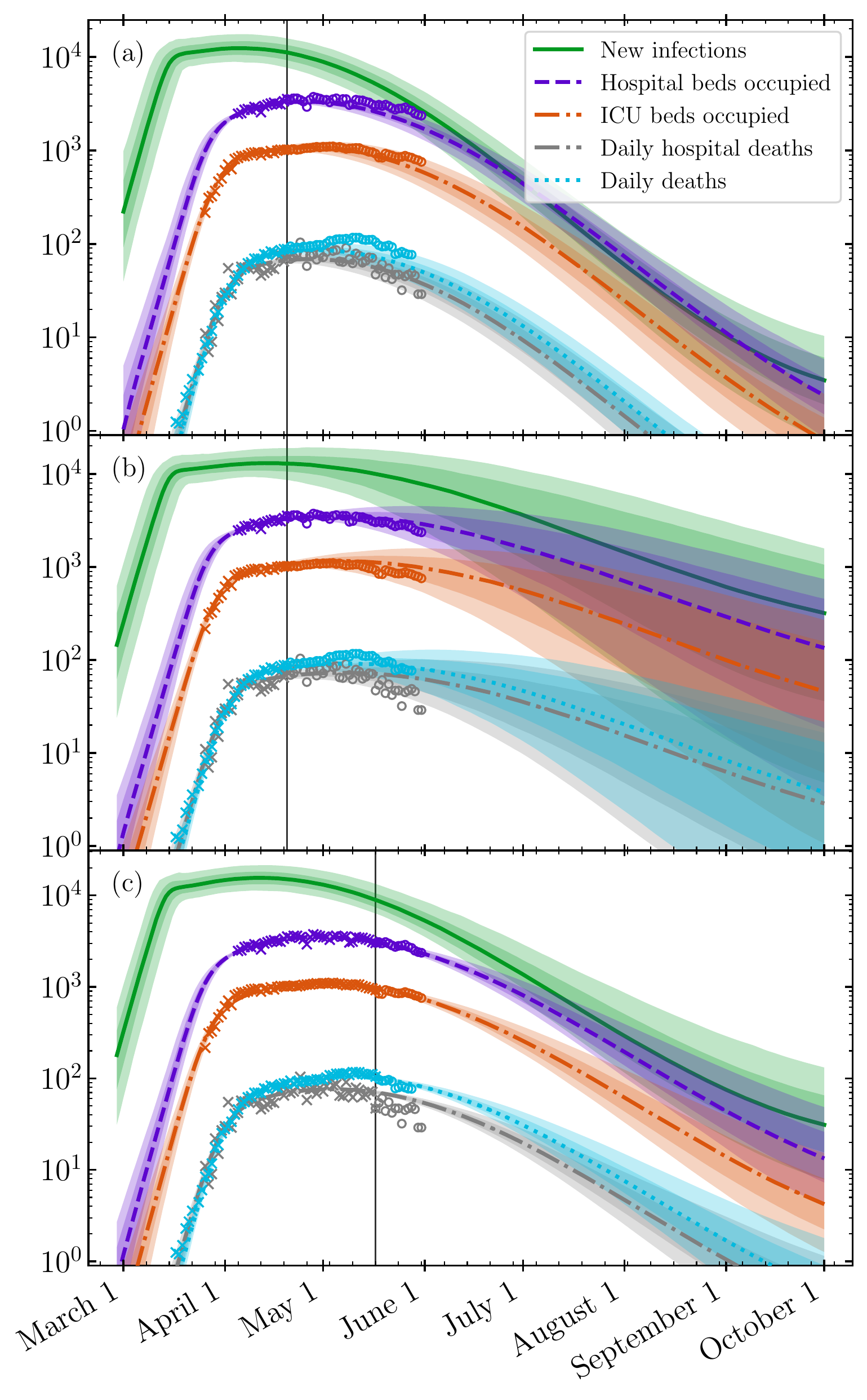}
\caption{
    Fits and predictions of our model of the entire state of Illinois under the continuation of
    the Stay-at-Home order and social distancing measures,
    resulting from parameter inference which
    (a) fixes $A_\mathrm{SF} = 0.2$ and calibrates to data through April 20, 2020,
    (b) samples $A_\mathrm{SF}$ and calibrates to data through April 20, 2020,
    and (c) samples $A_\mathrm{SF}$ and calibrates to data through May 17, 2020.
    In each case, the data used for calibration is denoted by crosses,
    while data from later dates (open circles)
    allow a comparison of model predictions with subsequent real world observations.
    Each panel's vertical line marks the end of data used for calibration.
    Solid curves denote median model predictions at a given time,
    while the shaded regions denote the 68.4\% (darker shading) and 95.6\% (lighter shading)
    confidence intervals of a particular output.
    We depict the daily incidence of new infections (green, solid),
    the number of occupied hospital beds by confirmed or tentative
    COVID-19 patients (purple, dashed),
    the number of ICU beds occupied by confirmed COVID-19 patients (orange, dot-dashed),
    the number of hospital deaths by confirmed COVID-19 patients (grey, dot-dashed),
    and total daily deaths of COVID-19 patients (light blue, dotted).
    We remove reporting artifacts by plotting daily death data smoothed by a 7-day running average.
}
\label{fig2}
\end{figure}

In Fig.~\ref{fig2} we show the fits and predictions of our model
for the entire state of Illinois assuming that once implemented,
the state-imposed and self-regulated social distancing behavior of the population
maintains the same level until the end of the simulation.
We report the median and the 68.4\% and 95.6\% confidence intervals of several
dynamical outputs of our model, obtained from an ensemble of forward simulations
that sample the posterior distribution over model parameters.
Fig.~\ref{fig2} presents three separate calibrations: (a)
using data up through April 20, 2020 and assuming a fixed seasonal forcing
amplitude $A_\mathrm{SF} = 0.2$, (b) again using data through April 20, 2020
but instead sampling over $A_\mathrm{SF}$, and (c) using data through
May 17, 2020 and sampling over $A_\mathrm{SF}$.

First, comparing panels (a) and (b) demonstrates that a 20\% seasonal forcing effect
produces a worse projection and may be an overestimate; while our model does not infer the origin
of any yearly, periodic modulation to $R_t$ (or lack thereof), by mid May the forecasts
disagree with the data.
By contrast, the uncertainty introduced by sampling $A_\mathrm{SF}$ produces a more
robust fit to the long plateau exhibited in all data sources.
Next, comparing panels (b) and (c) shows that while short-term predictions are
broadly consistent, the spread of model forecasts become narrower.
In particular, including data up to May 17, as in panel (c), enables the model to identify that the plateau
is beginning to bend, in contrast to the model shown in panel (b), in which a continued plateau is not precluded.
We investigate the predictive power of our model and calibration procedure in the next
subsection.

Hospital and ICU occupancy as well as deaths related to COVID-19 exhibit a long
plateau spanning from the beginning of April through at least mid May.
This behavior is not just due to the fact that mitigation reduced $R_t$ to almost
exactly one, but also because of the variance in when the
lagging indicators respond to the rate of infection.
Namely, while the various non-pharmaceutical interventions (NPIs) implemented in Illinois occurred on relatively short timescales
(as can be seen in the sudden change of slope for the daily incidence of new infections in Fig.~\ref{fig2}),
the variance in the delays between when individuals become, e.g., symptomatic and then hospitalized, introduces spread in the indicators' responses to changes in $R_t$.
As an example, some portion of individuals infected before any mitigation takes place will continue to
be admitted to the hospital well after mitigation occurs.
Indeed, this variability compounds with subsequent transitions, such that
daily deaths, being the final indicator, exhibit the most gradual change in incidence rate.
Our model's generality to arbitrary delay distributions makes it particularly well-suited
to accurately capture this effect.

\subsection{Robustness of predictions}

\begin{figure}[ht]
\centering
\includegraphics[width=\columnwidth]{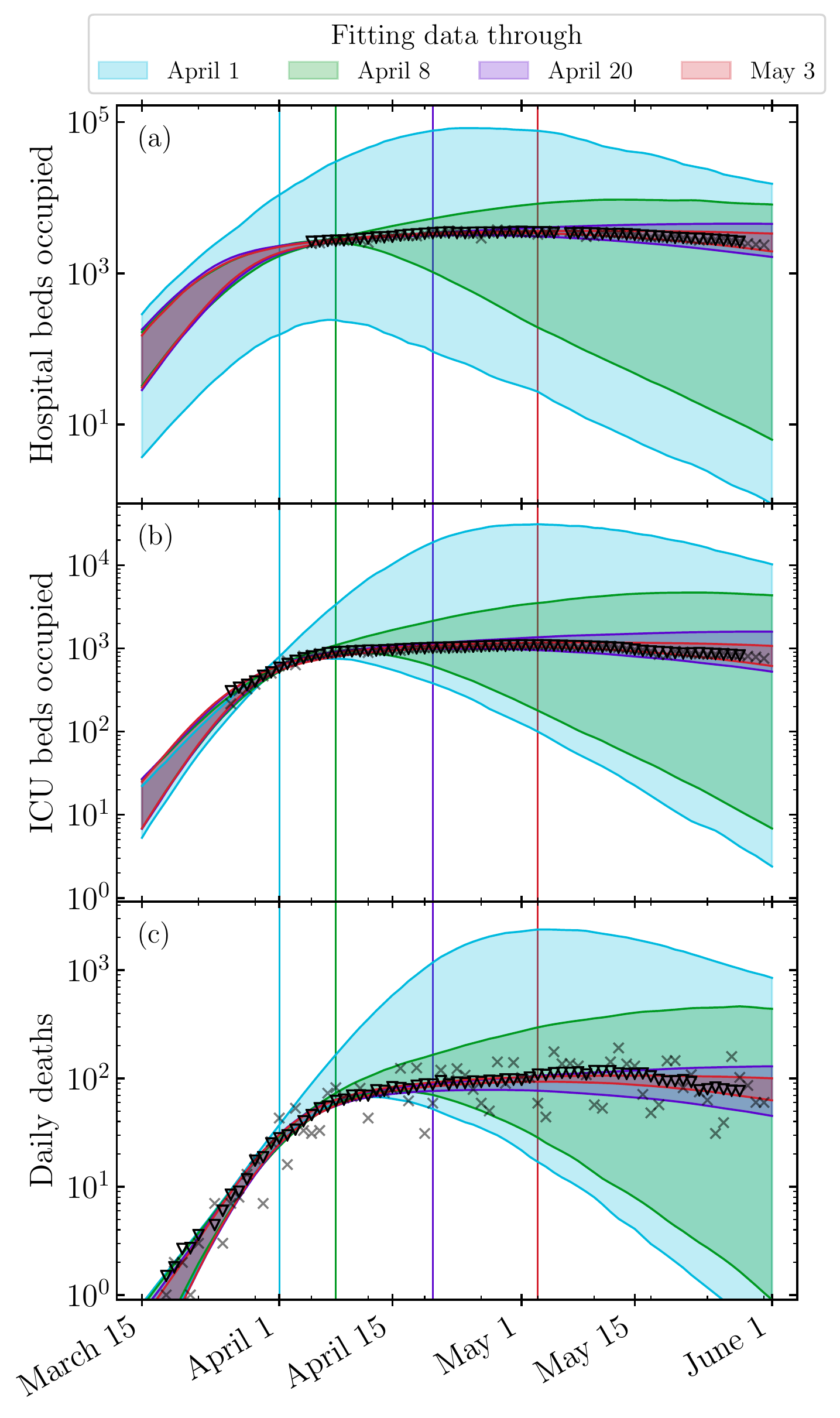}
\caption{
    Robustness of simulations for the COVID-19 epidemic in
    Illinois, evaluated by comparing 95.6\% confidence regions of predictions
    for parameter inference calibrated to data through April 1 (light blue),
    April 8 (green), April 20 (purple), and May 3, 2020 (red).
    The panels depict the number of hospital beds occupied (a), the number
    of ICU beds occupied (b), and the total daily deaths (c).
    In each panel, triangles depict the data smoothed by a 7-day rolling average
    and grey crosses the actual data.
}
\label{fig:robust}
\end{figure}

In order to explore the predictive capabilities of the model, we present
a series of benchmark simulations to compare the predictions
of models calibrated with increasingly recent data.
In Fig.~\ref{fig:robust}, we show
the fits to hospital beds occupied, ICU beds occupied, daily
hospital deaths, and daily total deaths in the entire state of
Illinois, calibrating with data up to April 1, 2020, April 8, 2020, April
20, 2020, and May 3, 2020.  During the time period studied here, the
Illinois Stay-at-Home order was issued, leading to an end to the
exponentially growing phase of the epidemic and a flattening of the
curve.
Due to transition delays, the quantities to which we calibrate exhibit exponential growth
as late as early April; for example, the
number of daily deaths does not flatten until around April 10, 2020.
Thus, the above-specified dates of calibration provide a test that
measures the ability of our model to anticipate the bending of the curve.

As expected, simulations with earlier terminal calibration dates
generate forecasts with larger, less restrictive confidence intervals:
with data from only the earliest stages of the epidemic, neither
continued exponential growth nor successful suppression of infections
due to mitigation can be ruled out. Between April 1 and April 8,
however, ICU occupation stopped growing exponentially, thereby
providing the first indicator by which the model can infer the effect
of mitigation. Indeed, even the April 8 forecast anticipates the
subsequent flattening in the daily deaths curve. Later forecasts (e.g.,
that of April 20) remain consistent with the April 8 model, while also
favoring a continuation of the plateau over a more rapid decline in use
of hospital resources and deaths. Finally, the May 3 model is largely
consistent with the April 20 one, but begins to project a slight
decline that agrees with the new data. This latest forecast should not
be expected to lie strictly within the confidence interval of the
previous, as any future changes in mitigation cannot be anticipated.

We also point out that the April 1 forecast for hospital occupation
spans several orders of magnitude on any given day (as seen in panel
(a) of Fig.~\ref{fig:robust}). This is due in part to the lack of
hospitalization data that could provide direct constraints, but is also
an inevitable feature of a system which exhibits exponential growth dynamics.
Furthermore, we hypothesize that the lack of hospitalization data
before the beginning of the plateau is at least partially responsible
for the April 1 forecast for ICU occupation and daily deaths being
relatively unconstrained. Since hospitalizations serves as the earliest
available indicator for the progression of the epidemic, we hypothesize
that if it were supplied with this data before April 1, the model would
have been able to discern that the data were no longer consistent with
an exponentially-growing epidemic. This observation underscores the
importance of rapid and reliable reporting of hospitalization data---or
even better, robust and representative testing for positive cases--- in
the context of modeling epidemic dynamics.

The disappointingly short horizon of predictability for epidemic models
when $R_t > 1$ shown here represents a fundamental limitation of
forecasting, in much the same way that chaotic dynamics limits weather
prediction, and this issue has been noticed in other independent studies
\cite{korolev2020identification,castro2020predictability,roda2020difficult}.  However,
the exponential sensitivity has a silver lining: small changes to
transmission can have large impacts on the overall trajectory and
fatality of the disease.

In summary, the model curves fitted after April 8, 2020, i.e., with
data from the plateau, follow the trends of the data well. We conclude
that the model can be characterized as semi-quantitative and that it is
capable of capturing broad epidemic dynamics and fitting empirical
data. In this sense, it can serve as a useful tool to make short term
predictions that may be useful for planning purposes.

\subsection{Regional modeling}

To account for differences in the epidemic in different parts in the
state, we simulate the epidemic trajectory in the four distinct Restore
Illinois regions separately. Each of these ``super-regions'' is
composed of multiple of the Emergency Medical Services (EMS) regions
defined by the Illinois Department of Public Health (IDPH)
~\cite{EMS_map}. The Northeast super-region includes the city of
Chicago and its suburbs, EMS regions 7 through 11. The North-Central
super-region contains EMS regions 1 and 2, the Central super-region EMS
regions 3 and 6, and finally the Southern super-region comprises EMS
regions 4 and 5. In the absence of detailed transportation data, we
assume no population transfer between these super-regions, and so
unlike, e.g., the model for the state of Georgia in
Ref.~\cite{weitz2020} and Italy~\cite{gatto2020spread}, our regional
modeling is not a genuine metapopulation model for the state.

\begin{figure*}[ht]
\centering
\includegraphics[width=.8\textwidth]{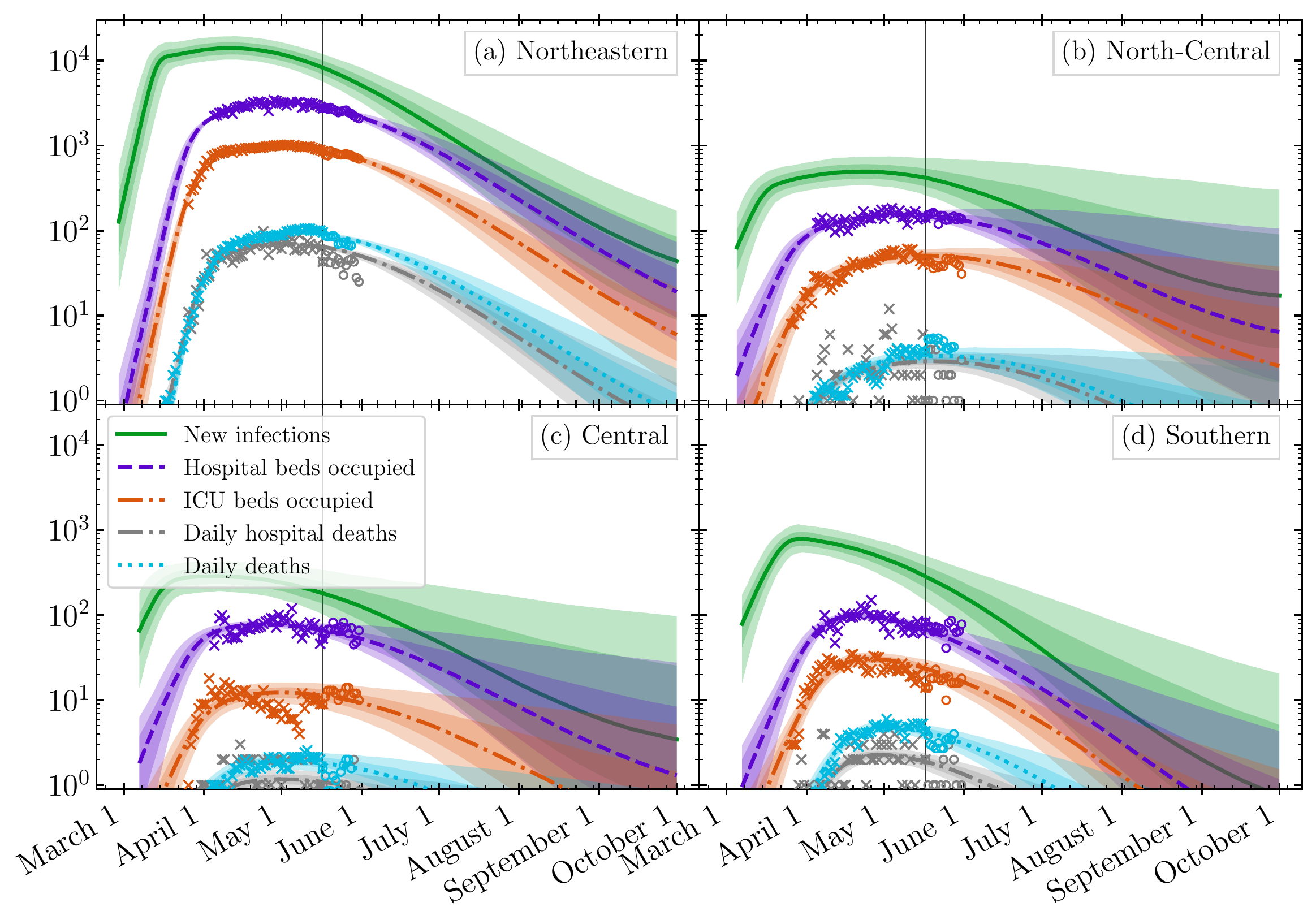}
\caption{
    Fits and predictions of our model under the baseline scenario
    with the Stay-at-Home order and social distancing maintained
    until October 1, 2020. The model is calibrated with data (crosses) through
    April 20, 2020 separately for each of the four super-regions of the state:
    (a) North-Central, (b) Northeast, (c) Central and (d) Southern.
    The data from later dates (open circles) were not used in our fits
    and allow a comparison of model predictions with real world observations.
    Solid lines of different denote the median model prediction at a given time.
    The shaded regions denote the 68.4\% (darker shading) and 95.6\% (lighter shading)
    confidence intervals, obtained as quantiles of an ensemble of forward simulations
    which sample the posterior distribution over model parameters.
    We depict the daily incidence of new infections (green, solid),
    the number of occupied hospital beds by confirmed or tentative
    COVID-19 patients (purple, dashed),
    the number of ICU beds occupied by confirmed COVID-19 patients (orange, dot-dashed),
    and total daily deaths of COVID-19 patients (light blue, dotted).
    We remove reporting artifacts by plotting daily death data smoothed by a 7-day running average.
}
\label{fig3}
\end{figure*}

In Fig.~\ref{fig3} we show the fits and predictions of our model
calibrated to data up to May 17, 2020 for each of these four regions.
We report inferred model parameters in Table~\ref{table:parameters1}.
Our median estimates of the basic reproduction number $R_0$ at the
start of the epidemic are consistently above 1, ranging from $2.4 \pm
0.16$ for the Northeastern region including Chicago and its
suburbs to $1.7 \pm 0.16$ for the Central region including
the University of Illinois at Urbana-Champaign.

The per capita daily deaths and illnesses is at least ten-fold higher in Chicago and its suburbs compared to the downstate areas of Illinois.
This is likely due to increased contact density in the upstate region, reflected by a higher initial $R_0 \approx 2.3$ compared to $\approx 1.8$ in the downstate regions, coupled with the fact that mitigation began earlier relative to the start of the epidemic in some downstate regions.
In regions where the virus entered the population later, the epidemic had a shorter phase of unmitigated exponential growth.

The values of $R_t$ corresponding to the post-mitigation basic
reproduction number are very close to $1$, reflecting the approximately
constant number of hospital and ICU beds occupied by COVID-19 patients
and daily deaths from COVID-19 in different super-regions in Illinois
during much of April and May 2020.

\subsection{Evaluation of parameter fits}\label{sec:evaluation-params}

\begin{figure}[ht]
\centering
\includegraphics[width=\columnwidth]{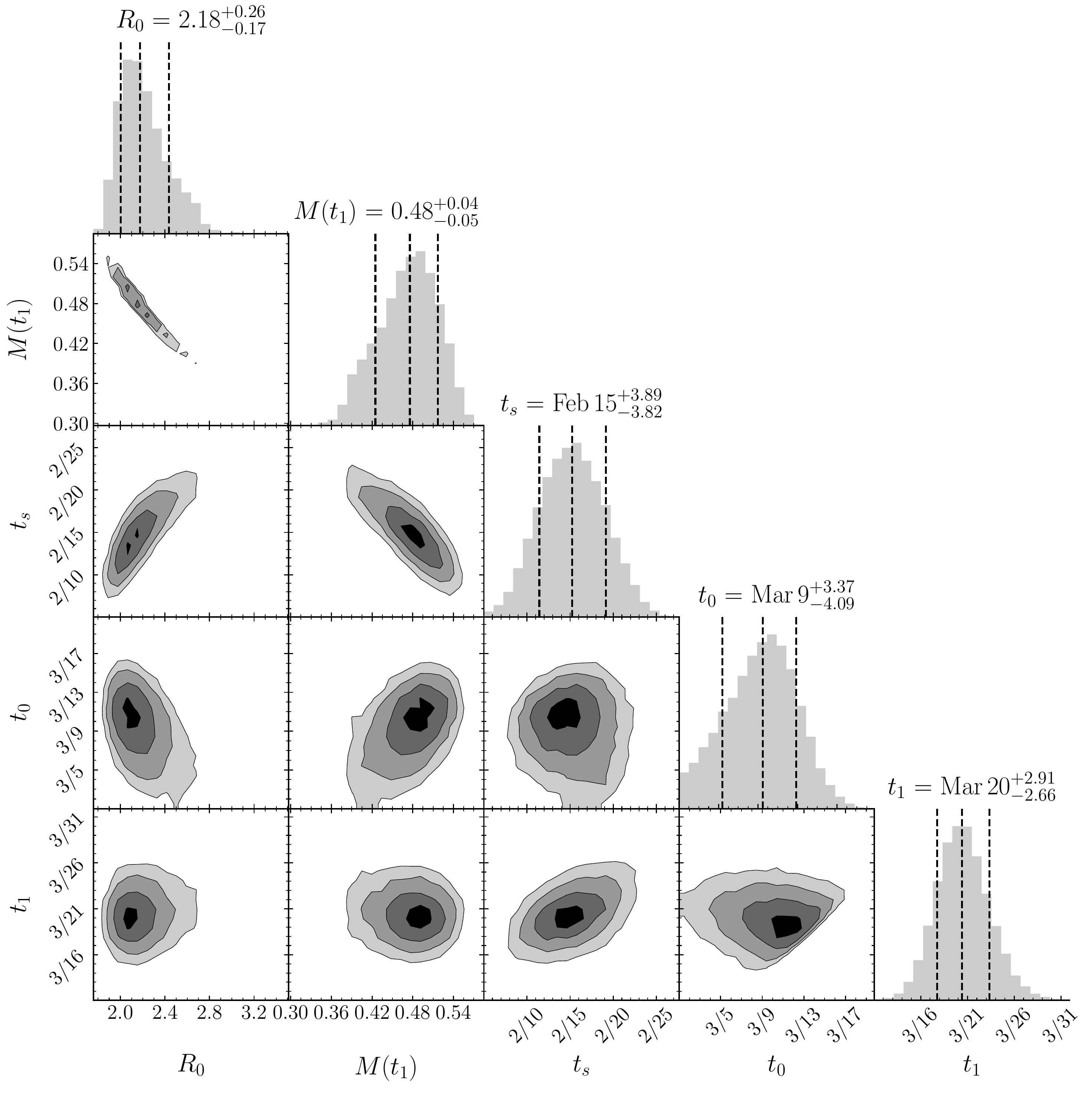}
\caption{
    Joint posterior distributions of pairs of the main parameters
    of our model fitted to the all-state data in Illinois up to May 17, 2020. The
    variables shown are the initial value of the reproduction number, $R_0$,
    the mitigation factor $M(t_1)$,
    the start date of the epidemic $t_s$, the day the mitigation factor
    begins to deviate from 1, $t_0$,
    and the day the mitigation reaches its asymptotic value, $t_1$.
}
\label{fig:corner}
\end{figure}

In Fig.~\ref{fig:corner}, we show a subset of the joint posterior probability distributions
for model parameters relevant to the parameterization of the mitigation factor $M(t)$ as specified above,
fitting to data shown in Fig.~\ref{fig2}.
The correlations between some pairs of fitted parameters, e.g., between $R_0$
and the start date of the epidemic $t_s$, are
reflected in the ellipsoidal shape of the posteriors. This is sensible: the later
the epidemic begins, the larger the basic reproduction number must be in order
to fit the data.

In Table~\ref{table:parameters1}, we report the parameter values our model infers when fitting to data for different regions of the state and over different time ranges. We also report the effective reproduction numbers $R_t$ as evaluated on May 1, 2020.
On May 1, 2020, $R_t$ appears to have barely dropped below unity, suggesting that mitigation efforts may have only marginally halted the exponential growth of the epidemic at that time.

\subsection{Comparison with mobility data}
\label{sec:mobility_comparison}

\begin{figure}[ht]
\centering
\includegraphics[width=\columnwidth]{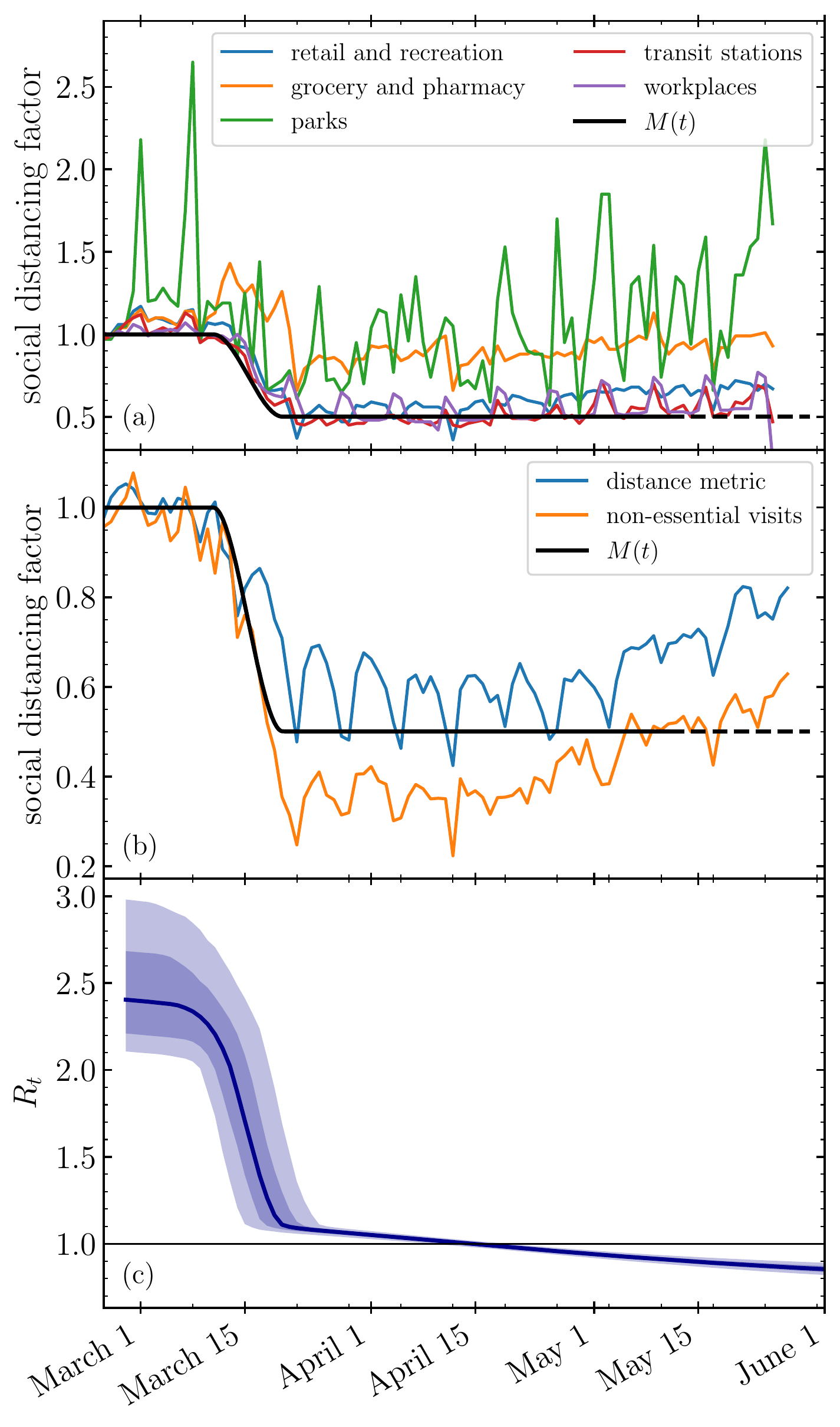}
\caption{
    The time-dependent mitigation factor $M(t)$ defined in Section~\ref{subsec:model-time-description}
    as inferred from model fits
    compared to measures of mobility in Illinois provided by Google (a) and Unacast (b).
    In (c), we also plot the inferred evolution of $R_t$ as defined in Eq.~\ref{eqn:Rt}, which we draw from a calibration to data from the entire state of Illinois through May 17, 2020.
    Shaded regions denote the 68.4\% (lighter shading) and 95.6\% (darker shading)
    confidence intervals, while the solid curve denotes the median.
}
\label{fig:mobility}
\end{figure}

While our model does not provide a microscopic description of social interactions and
movement in the population, we may
evaluate our fitted $M(t)$ relative to social mobility indices
derived from cell phone data~\cite{Goog,Unacast}.
In the top panel of Fig.~\ref{fig:mobility}, we plot the time dependence of several
mobility indices reported by Google~\cite{Goog} for the entire state of
Illinois, measuring change in visits to destination points categorized
as retail and recreation, grocery and pharmacy,
parks, transit stations, and workplaces.
The Unacast mobility data~\cite{Unacast}
is shown in bottom panel of Fig.~\ref{fig:mobility},
depicting an effective distancing metric and a measure of trips to so-called non-essential
destinations.

Remarkably, although the model is supplied with no prior information on non-pharmaceutical interventions,
the inferred dates and magnitude of mitigation
agree with the start and end dates of the largest drop in both sets of mobility data.
Comparing to the Google mobility data, our $M(t)$ curve exhibits an amplitude,
start date and end date consistent with indices corresponding to retail and recreation, transit stations and workplaces.
On the other hand, parks and grocery and pharmacy show a more modest reduction which
still matches the time frame of $M(t)$.
In the Unacast data, both metrics also appear to match the time-dependent change in $M(t)$.
Note that both datasets evince increased movement at later dates.
Because we here parameterize mitigation as a single transition,
our fits to $M(t)$ would not reproduce this recent increasing trend.

Our procedure is distinct compared to several previous efforts to incorporate mobility into models of COVID-19 dynamics,
which either impose that changes in transmission coincide with known dates of non-pharmaceutical interventions,
or scale $R_t$ according to reductions in mobility~\cite{flaxman2020report, Juliette2020report}.
The fact that our model independently recovers a trend in mitigation consistent with
mobility measures speaks to
its flexibility and calibration procedure.

\section{What-if scenarios}

We now consider possible future scenarios and alternative historical scenarios
in which non-pharmaceutical interventions are lifted or were never implemented at all.
The former enables a model-based assessment of the risk of, e.g., lifting Stay-at-Home orders
on a certain date, while the latter demonstrates the impact that previously-implemented measures
have already had on outcomes.
Our analysis focuses on two key measures for guiding and justifying policy decisions:
the stress imposed on the healthcare system and the death toll.

\subsection{No Stay-at-Home order}

\begin{figure}[ht]
\centering
\includegraphics[width=\columnwidth]{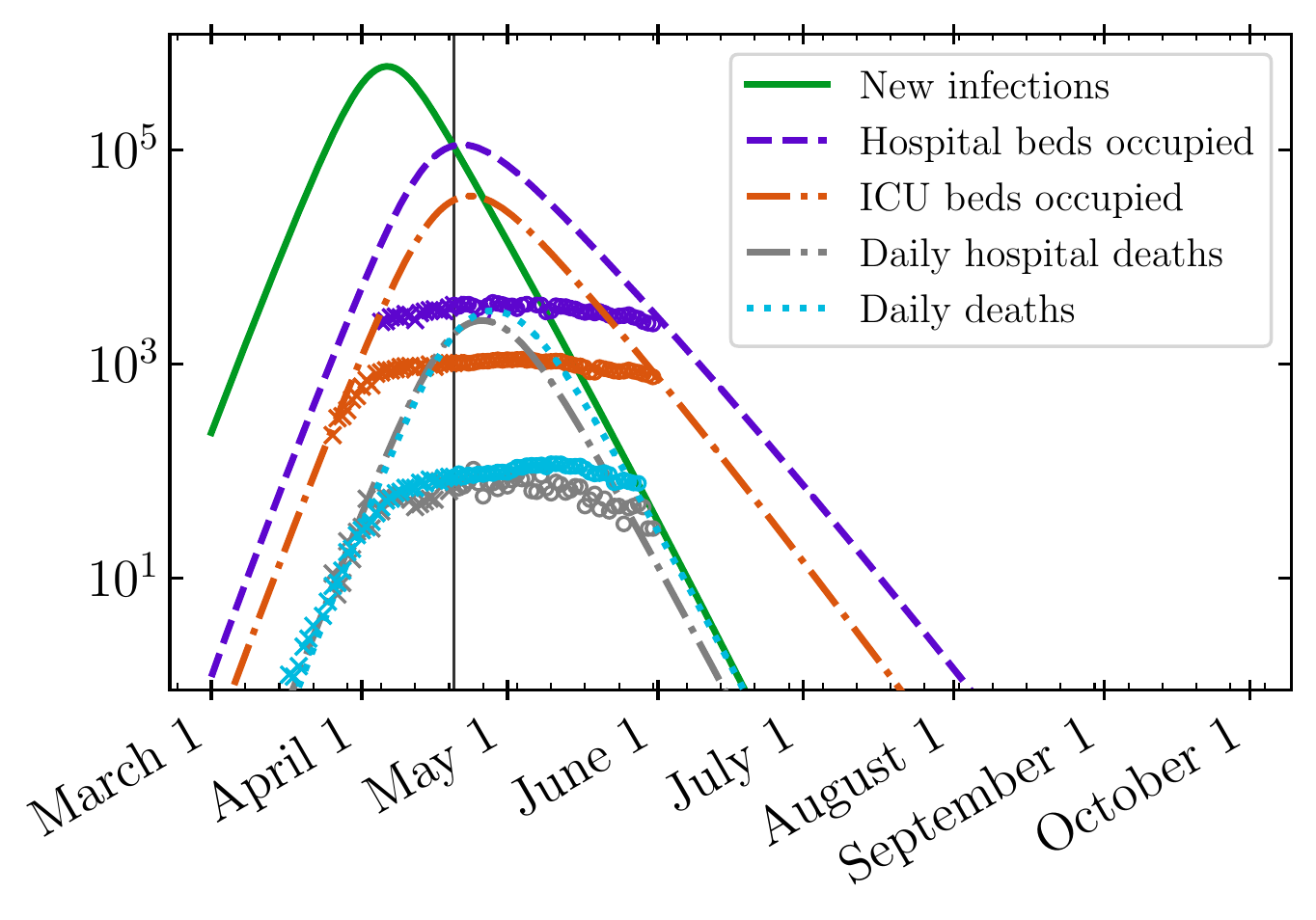}
\caption{
    Counterfactual simulation of the COVID-19 epidemic in the absence
    of government or population self-imposed social distancing measures.
    The model is calibrated to data through April 20, 2020 (crosses); we then
    artificially set $M(t_1) = 1$, i.e., assume no mitigation ever takes place.
    The data from later dates (open circles) are not used in our fits
    and demonstrate the effect that real-world NPI strategies had on the epidemic.
    We depict the daily incidence of new infections (green, solid),
    the number of occupied hospital beds by confirmed or tentative
    COVID-19 patients (purple, dashed),
    the number of ICU beds occupied by confirmed COVID-19 patients (orange, dot-dashed),
    and total daily deaths of COVID-19 patients (light blue, dotted).
    We remove reporting artifacts by plotting daily death data smoothed by a 7-day running average.
}
\label{fig:neversip}
\end{figure}

We first investigate the trajectory of the COVID-19 epidemic in
Illinois in the absence of any form of social distancing or mitigation
measures, whether self-imposed or mandated by the government.
We conclude that rate of hospital-bound deaths,
ICU bed occupancy, and hospital bed occupancy would be higher than what
actually took place by an order of magnitude or more, as shown in Fig.~\ref{fig:neversip}.
The Stay-at-Home order and self-imposed social distancing measures
were clearly crucial to flattening the curve.

In Ref.~\cite{Maslov2020} we made an early estimate of the ICU
utilization in the city of Chicago under two scenarios: one in which
the Stay-at-Home order was issued on March 20, 2020, and another in
which the order was delayed by 20 days. Under the first scenario, the
ICU utilization by COVID-19 patients never exceeded the number of ICU
beds not occupied by other patients, while under the second scenario it
exceeded ICU capacity by nearly a factor of ten. This example
highlights the cost of mistiming in NPIs~\cite{morris2020optimal}.  In
spite of inevitable uncertainties associated with these early
estimates, that work correctly identified the timing of the peak in ICU
demand to happen on or around April 22, 2020.  However, the magnitude
of this peak was underestimated in this study, because the scenario
assumed  that post-mitigation value of $R_t=0.9$ would be achieved by
social distancing. In reality, the response of the population to the
Stay-at-Home order in Chicago was somewhat weaker resulting in a larger
value of $R_t$ and about a three-fold higher peak ICU bed occupancy
than we had predicted.

\subsection{Partial removal of Stay-at-Home order}

\begin{figure}[ht]
\centering
\includegraphics[width=\columnwidth]{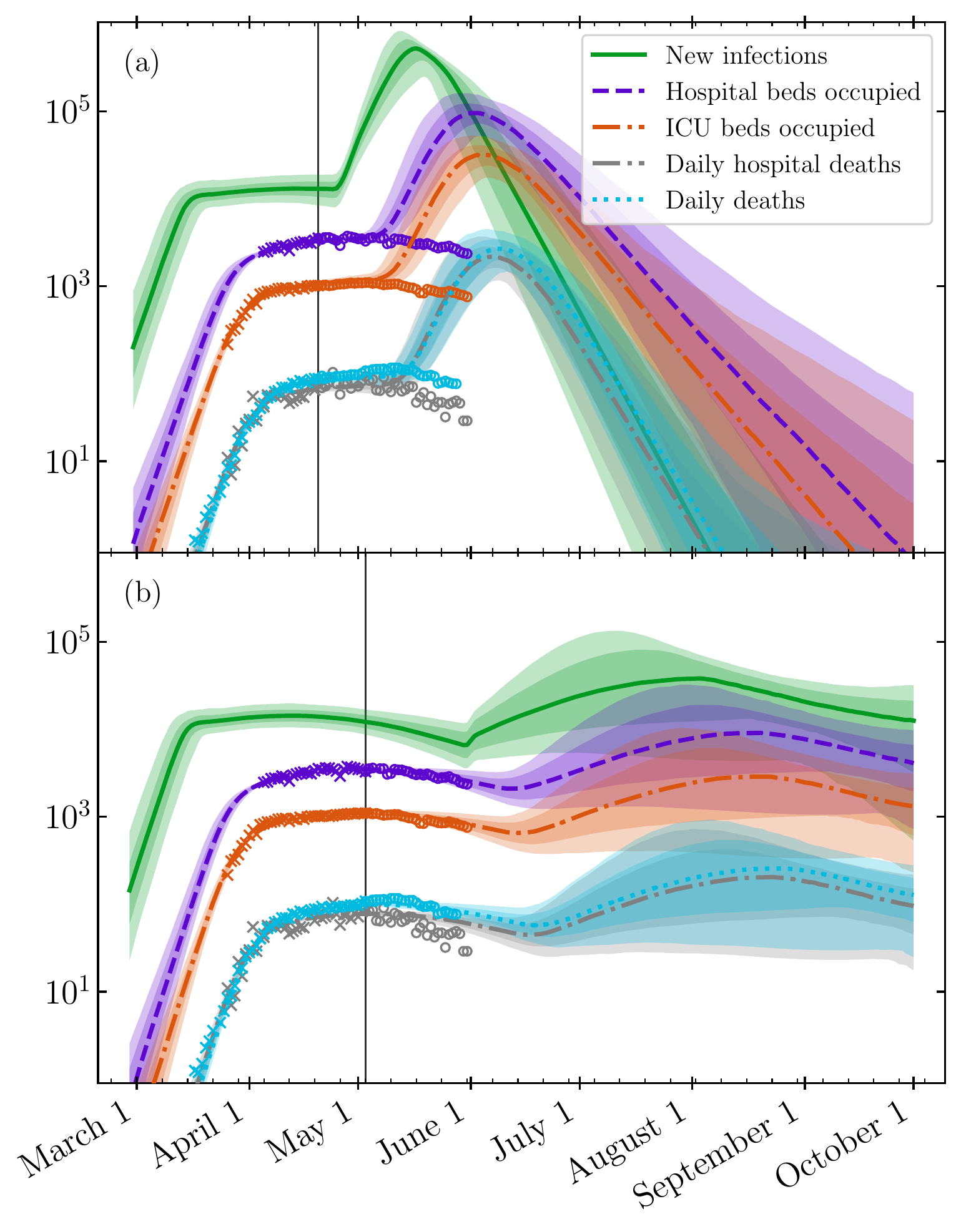}
\caption{
    Simulation of different scenarios for relaxation of the Stay-at-Home order
    for the entire state of Illinois.
    Each panel's vertical line marks the end of data used for calibration.
    We display two scenarios: (a) where the mitigation actor $M(t)$ returns to 1 on April 24, 2020,
    corresponding to a complete lack of social distancing,
    and (b) where on June 1, 2020 the effect of NPIs is reduced by 30\%.
}
\label{fig:endsip}
\end{figure}

The state of Illinois lifted its original Stay-at-Home order on May 30, 2020~\cite{Illinoisresponsewiki}.
In Fig.~\ref{fig:endsip} we consider two scenarios for the lifting
of Stay-at-Home orders for the entire state of Illinois.
In the first scenario, mitigation effects are completely removed and $M(t) = 1$
for times $t$ after June 1, 2020.
We also consider the more conservative case that mitigation recovers by 30\% to
$M(t) = M(t_1) + 0.3 (1 - M(t_1))$ for $t$ after June 1, 2020.
This second scenario assumes that a combination of self-regulation and remaining
government-imposed mitigation measures, such requiring wearing masks, banning
large gatherings, etc.,~results in only a partial reduction of the effective
mitigation factor.

The first scenario exhibits a substantial
second wave, with rapid and strong peaks in all quantities occurring
successively through the month of July.
In the second, weaker (and perhaps more realistic)
scenario, a second wave still occurs but does so later and with a reduced peak height.
In this case, the peak demand for hospital and ICU beds and number of deaths are reduced.
\begin{figure*}
\centering
\includegraphics[width=.8\textwidth]{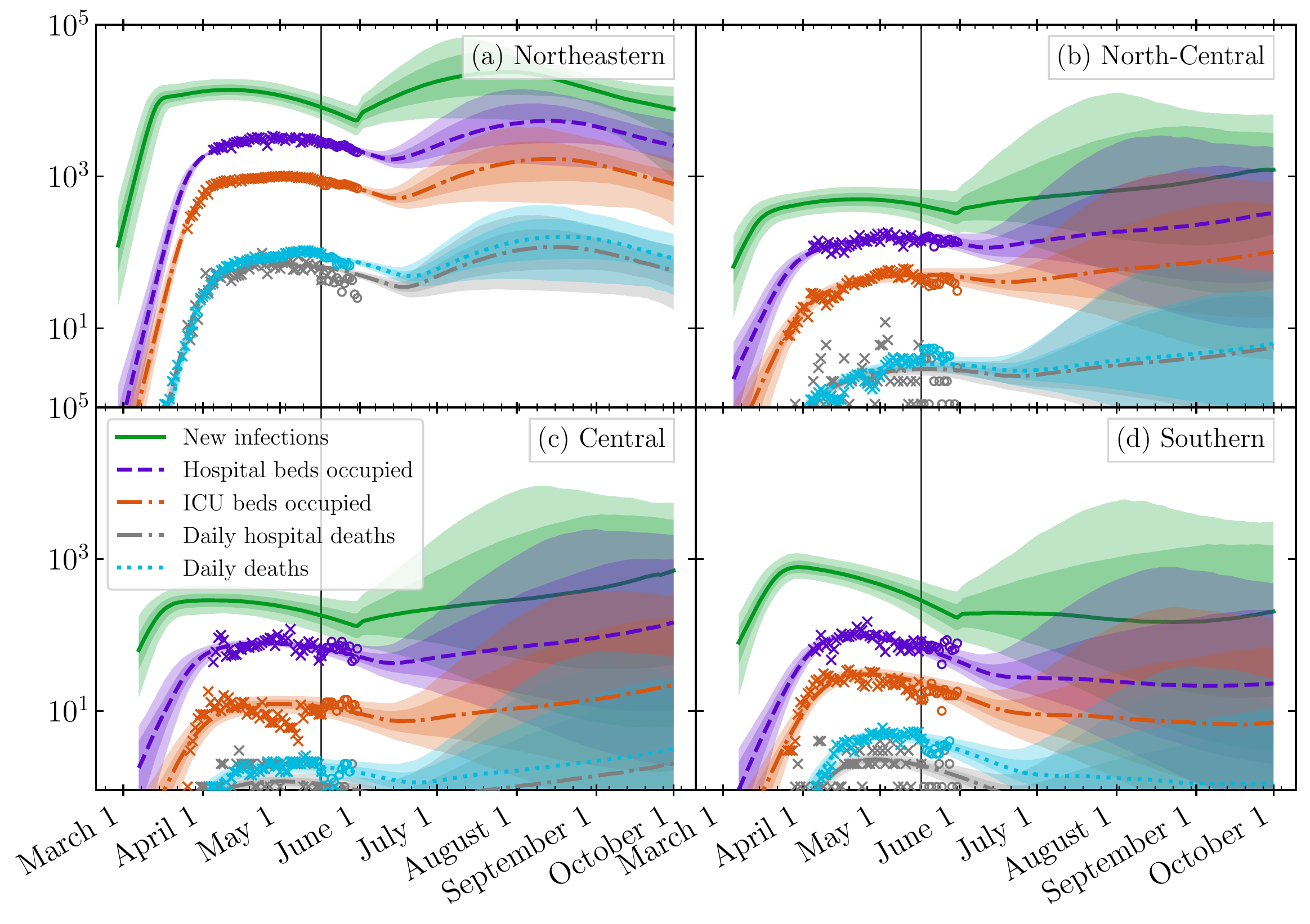}
\caption{
    Fits and predictions of our model under the a partial release of the
    Stay-at-Home order on June 1, 2020 to a mitigation factor
    $M(t) = M(t_1) + 0.3  \times \left(1 - M(t_1)\right)$.
    The panels and and plots denote the same regions and data sources as those in
    Fig.~\ref{fig3}.
}
\label{fig:partial_endsip_super_regions}
\end{figure*}

An aggregated model of the entire state does not describe its
heterogeneous population structure, which may be particularly important
in forecasting beyond the lifting of the Stay-at-Home order.
We now report separately the results of modeling
the expiration of the Stay-at-Home (with a 30\% reduction in mitigation)
in each of the four aforementioned super-regions, with the caveat that
we are unable to take into account possible transfers of people
between the regions.
As in Fig.~\ref{fig:partial_endsip_super_regions}, we first calibrate models
using data through May 17, 2020, imposing again that mitigation recovers by 30\% to
$M(t) = M(t_1) + 0.3 (1 - M(t_1))$ for $t$ after June 1, 2020.

In each super-region, a second wave is clearly visible, as all four populations
still today have a significant number of infectious people at large.
Although the Northeast region, which includes Cook County and metropolitan Chicago,
has the biggest second wave in absolute numbers,
its relative impact on the epidemic trajectory is smaller merely
because the epidemic has not declined to the same extent as the other three regions.
Under the worst case scenario, the more sparsely populated Southern, Central, and North-Central regions have relatively larger second waves because the epidemic as of yet is relatively less active in those regions compared to the Northeastern region, which includes Chicago and its suburbs.

We note that the magnitude of the second wave may be containable if rapid and efficient
testing, contact tracing, and isolation mitigation strategies are
employed~\cite{hellewell2020feasibility,Ferretti2020}.
Following standard protocols~\cite{GWU}, we estimate that even if purely
manual contact tracing is employed for all three stages of tracing
(case identification, tracing, and follow-up), the number of contact
tracers required is approximately $8.3$ times the daily number of new cases identified.
A considerable reduction in workforce and an increase in efficiency can be obtained by electronic measures.
Nevertheless, the potential magnitude of the second wave in this scenario suggests that contact
tracing will be challenging and require extraordinary resources to execute.

\section{Discussion}

Modeling plays an important role in the societal response to the
COVID-19 pandemic, and a variety of approaches are used to inform public
health policy, guide resource allocation,
and plan non-pharmaceutical
interventions~\cite{Holmdahl2020}.  The present study of the spread of
COVID-19 in Illinois reveals both the strengths and limitations of
modeling, and provides potentially actionable insights into the
future spread of the disease.
We begin with some technical points and
best practices that we have developed during our work.

\medskip
\noindent
\textit{Model calibration:} Our analysis highlights the
importance of choosing appropriate data with which to calibrate models,
and to perform calibration with precision.  Due to the large number of
parameters that inevitably enter epidemiological models, calibration
requires parameter inference in a high dimensional space
with strong potential for improper fits resulting from failure to reach
global optima.
Although the MCMC methods we use are
computationally intensive, they are relatively efficient in exploring
high dimensional, multi-modal distributions,
and converge to well-behaved global posterior probability distributions.
Bayesian inference enables the incorporation of previous studies
(e.g., meta-analyses) to provide reasonable priors on
parameters which are poorly constrained by the available data.
As an example, although the data we calibrate to does not constrain the prevalence
of the infection, we systematically account for this uncertainty by informing our
prior on the infection-fatality rate (IFR) from serological studies~\cite{systematic_review_ifr}.
The IFR is an important variable in terms of
disease outcomes, and so model predictions must systematically account for
the uncertainty in this variable.

In our analysis, we have taken data at its face value. A more thorough
analysis could account for differences in the trustworthiness of data,
e.g., programmatically dealing with uncertainties associated with
classification of early hospital admissions as either COVID-positive or
under investigation. Such an approach might also be able to consider
the number of individuals who test positive for the virus. Furthermore,
because we do not explicitly treat underreporting, data from early in
the epidemic may be biased low.

\medskip
\noindent
\textit{Error estimation:} The MCMC procedure generates a complex, high
dimensional probability distribution that may be sampled to estimate
future epidemic trajectories.
While one might be tempted to simply generate
the maximum likelihood trajectory for forward (in time) simulation,
this can be misleading.
The probability distribution of trajectories may have a maximum likelihood
trajectory that is favored only slightly above other trajectories yet
is noticeably different from the mean or median estimator.
We thus sample trajectories forward in
time, and at each time point we evaluate the median and
confidence intervals of outputs.
Although the curves we depict do not correspond
to a trajectory that would be realised in practice, together with confidence intervals, they provide a meaningful
description of the range of forecast results.

\medskip
\noindent
\textit{Robustness:} In order to ascertain the predictive
power of our model, we assessed to what extent it was able to make
predictions beyond the range of data to which it had been
calibrated.  We first considered a rather severe test: how far in advance can the
model predict the abrupt flattening of the epidemic curves that
occurred after April 10, 2020?
Unsurprisingly, the answer here was not particularly encouraging
because the curves dramatically switch from exponential
growth to a phase of much slower variation or plateau.
We were only able to fit daily death and ICU occupancy data before the flattening,
since hospitalization data was not available until mid-April.
As a result, none of the information feeding into
the model dynamics were able foreshadow the onset of the plateau.

Nevertheless, such self-consistency checks are important
validations of the modeling process, and should be attempted if there
are early enough data available.
Encouragingly, the range of our model
estimates decreased in a hierarchical and consistent manner, with projections from
earlier points in time bounding later projections.
Of course, it is not possible for our model to predict future changes to the strength of mitigation.
Furthermore, our mitigation model only implements a single event and thus cannot account for future short term changes.
Real world considerations, such as holidays and quarantine fatigue, would be inconsistent with our mitigation model approximation on long timescales.
As such, we do not necessarily expect future predictions (which could account for future changes
in mitigation) to lie within the confidence bounds of older ones.

\medskip
\noindent \textit{Correlation with mobility data:}
For policy and planning purposes, it is important to evaluate the
effectiveness of various measures in mitigating the epidemic relative to their, e.g.,
economic costs.
Although epidemiological modeling can in principle provide insight into the former issue,
doing so would require a more fundamental description than the one presented here, for example making use of
agent-based models~\cite{bansal2007individual,pastor2015epidemic,kim2018agent}
or employing spatially-structured and heterogeneous network descriptions~\cite{rock2014dynamics}.
Mean field models, which assume a well-mixed population, do not capture
the microscopic effects that could be correlated to specific mitigation strategies.

Nevertheless, mobility data present a potential means to guide and evaluate
which policies and social interventions
lead to the strongest reduction in disease transmission.
While our treatment does not provide such a description or analysis,
the correlation between the effective measure of the impact of NPIs, $M(t)$
(see Eq.~\ref{eqn:Rt}), and mobility data presented in Fig.~\ref{fig:mobility}
is striking. Note that in contrast to Ref.~\cite{Juliette2020report}, we do not use the mobility data as an input to our calculations.
Although our model makes no claims about the effect of changes
in the reported mobility indices, this observed correlation
encourages more detailed modeling to evaluate the impact of different social
distancing measures. Furthermore, our results support the utility of anonymized, aggregated mobility
data as a potential low-latency measure of the impact of, e.g., government-mandated
social distancing measures on actual population movement, especially since such indicators may
provide a
predictor of the measures' influence on epidemic dynamics.
A robust understanding of the mechanistic relationship between mobility measures
and the spread of the infection
could also guide efficient testing and contact tracing strategies,
for example using risk-based surveillance methods~\cite{stark2006concepts,foddai2020surveillance}.

\begin{figure}[ht]
\centering
\includegraphics[width=\columnwidth]{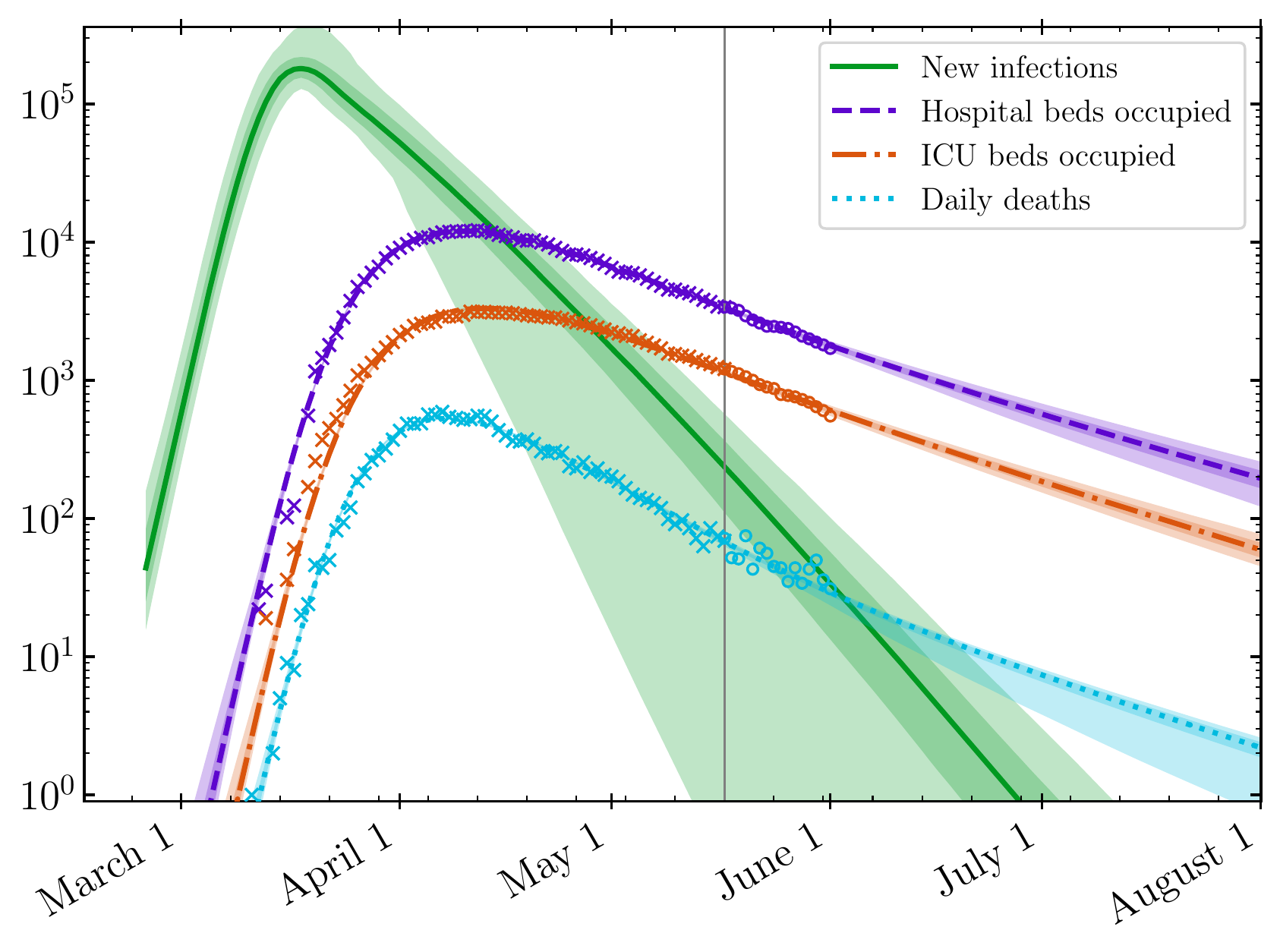}
\caption{
    Model fit and prediction for hospital and ICU bed occupancy~\cite{Thecityrepo,Thecityny} and deaths~\cite{Nychealth} in New York City using public data. Data from before the gray line on May 17, 2020 (crosses) are used for calibration.
    More recent data are plotted as circles.
    The model predictions shown here do not account for changes to mitigation after May 17, 2020.
}
\label{fig:nyc_example}
\end{figure}

\medskip
\noindent \textit{Model generality:} Although we designed and tested
our model using data for the state of Illinois, the model is general
and can be fit to other geographical regions as well, as long as there is
availability of hospital, ICU and death data. Using the procedure described
above, we calibrate model parameters to public data for hospital and
ICU bed occupancy and daily deaths in the New York City data published
by the New York City Department of Health and Mental
Hygiene~\cite{Nychealth,Thecityrepo,Thecityny}. We present the results
of the fit in Figure~\ref{fig:nyc_example}. With no modification, the
model generates a forecast that is remarkably consistent with
subsequent data. These fits are subject to the stated
limitations of our modeling.  In particular, future changes in
voluntary and state-mandated mitigation are of course not included.
Furthermore, there are possible indications of strong population
heterogeneity and overdispersion in the unexpected smallness of
$\tau_{\mathrm{disch}}$ compared to the value for Illinois.  A detailed
discussion of these effects is beyond the scope of this paper. We
report model parameter values in Table~\ref{table:nyc-parameters}.

\medskip
\noindent \textit{Spatial heterogeneity:}
In reality, the entire state of Illinois is not a single well-mixed system,
even if such approximations are frequently made~\cite{keeling2011modeling,rock2014dynamics,pastor2015epidemic}.
Illinois has a densely populated region in the northeast, and more
sparsely populated regions further west and to the south.
Our modeling identified that the status of the epidemic (and
the projections for scenarios in which mitigation is relaxed)
differ among the four super-regions.
In particular, in the Southern and Central regions,
the epidemic is declining more rapidly, due to a combination of
social distancing and contact tracing using pre-existing resources.
Modeling these regions separately, and including population transfer between them,
is essential to guide the implementation of region-dependent mitigation strategies
and to provide input to policy makers to prevent the resurgence of the epidemic.

\medskip
\noindent \textit{Relaxing mitigation:} Our simulations suggest
that it is too soon to lift all social distancing and mitigation
restrictions, as the significant number of currently-infectious
individuals would make a second wave inevitable.
The dynamics of a second wave and the approach to herd immunity
may be crucially sensitive to effects beyond a mean-field description.
For example, we do not account for super-spreader events which have played a major role in the spread of SARS and
MERS~\cite{LloydSmith2005,small2006super,kim2018agent} and are likely also
relevant for COVID-19~\cite{Liu2020}.
Similarly, we do not account for heterogeneities in the population structure.
Such features may accelerate the onset of herd immunity which in turn
decreases $R_t$. These differences have the potential to decrease the severity of the second wave and may enable more efficient containment.
We defer a treatment of these effects to future work.

\medskip
\noindent \textit{Additional limitations:} We end with a brief discussion
of additional limitations.
First, the importance of spatial structure,
heterogeneities in population susceptibility, and social network structure
are well-appreciated in the literature~\cite{bansal2007individual,keeling2011modeling,rock2014dynamics}.
We plan to investigate methods to model these effects and their impact on our results in future work.
In addition, important sources of error in our existing
model include the unknown impact of seasonal forcing,
the discreteness of populations,
and the effect of super-spreader events and behavioral response.

In our initial modeling~\cite{Maslov2020}, we used seasonal
forcing with $A_{\mathrm{SF}} = 0.2$ because similar effects had been documented for historical
coronaviruses~\cite{Neher2020}.
Because seasonal forcing varies slowly with time, it remains difficult to verify this assumption with the observables we consider.
Nevertheless, our parameter inference does not exhibit strong evidence for this level of seasonal forcing;
the role of climate in the early stages of epidemics is a question of
active investigation~\cite{Raines2020.05.23.20111278,Baker2020.04.03.20052787}.
Furthermore, non-zero values for $A_{\mathrm{SF}}$ do not necessarily ascribe an equivalent seasonal modulation to the disease's infectivity.

In addition, we wish to comment on the reported confidence intervals
obtained from parameter inference for both stochastic and deterministic
epidemic models~\cite{King2015}, because these can depend on whether
or not the model calibrations are performed for raw data or cumulative
data. Generally speaking the estimated intervals are
unrealistically small when models are fit to cumulative data
\cite{King2015}, with stochastic models giving slightly larger
intervals.  Our calculations are calibrated to raw data not
cumulative data, and so we do not expect the uncertainties in general
to be estimated inappropriately. Regardless, any deterministic model of
the epidemic trajectory will be inappropriate when few individuals are
infected, in spite of apparent certainty in posteriors.  Thus, in a
regime where the deterministic model is not even appropriate as a first
approximation, i.e., small populations,  the estimate of uncertainties
will not be a good indicator of the inapplicability of the
deterministic model. Moreover, it is well-known in ecology that the
discreteness of populations---the fact that individuals are quantized
and that birth-death processes are discrete---leads to important
qualitative phenomena ranging from population
cycles~\cite{mckane2005predator} to noise-induced pattern
formation~\cite{butler2009robust,karig2018stochastic}. These effects
are particularly important when numbers are small in the early and late
stages of the epidemic because our mean-field model is incapable of
representing the extinction state, i.e., when the number of infectious
individuals drops below one.  Once the epidemic is extinct in a
particular region, it can only re-emerge due to migration events, e.g.,
a super-spreader event like when university students return to campus.
In this regime, a prudent regional health department uses contact
tracing to contain outbreaks. Such mechanisms are not represented in
our modeling.

The third major limitation is that modeling an epidemic is very
different from modeling a physical system, even one as complex as a
weather pattern.  The transmission of an infectious disease involves a
collaboration between the virus and the host population: the host
population alters its behavior in response to its awareness of epidemic
progress, leading to policy steps that may increase or decrease
transmission, and self-regulation of social distancing by susceptible
and vulnerable populations.  Thus, it is important to emphasize that
predictions can easily be invalidated due to subsequent human actions
that cannot be anticipated, and will be impossible to model precisely.

\section{Conclusions}

In this paper, we have presented a mathematical model and computational
framework for recapitulating the COVID-19 epidemic. This model may be
used to infer values for the parameters that drive and represent the
progression of the disease. We use our calibrated model to provide
robust short-term forecasts of the epidemic trajectory in different
regions of the state and explore the effects that steps to relax social
distancing measures may have, especially in the context of a second
wave of the epidemic. The resulting highly-constrained and quantitative
narrative of the epidemic is a useful tool to inform scenarios for
sustainable monitoring and control of the epidemic.

\section*{Code availability}

The model and calibration framework described above have been
implemented in the open source Python 3~\cite{python} package
\href{https://pydemic.readthedocs.io/}{\textsf{pydemic}}. The source
code for \href{https://pydemic.readthedocs.io/}{\textsf{pydemic}} is
freely available online at
\url{https://github.com/uiuc-covid19-modeling/pydemic}. This work made
use of \textsf{NumPy}~\cite{numpy}, \textsf{SciPy}~\cite{scipy},
\textsf{pandas}~\cite{pandas}, \textsf{emcee}~\cite{emcee},
\textsf{corner.py}~\cite{cornerpy}, and \textsf{Matplotlib}
~\cite{matplotlib}.

\begin{acknowledgments}
We gratefully acknowledge discussions with David Ansell at Rush
University Hospital, Mark Johnson at Carle Hospital, Katie Gostic and
Sarah Cobey at University of Chicago, Jaline Gerardin at Northwestern
University, Charles Gammie at the University of Illinois, and Josh Speagle at Harvard University. The calculations we
have performed would have been impossible without the data kindly provided by
the Illinois Department of Public Health through a Data Use Agreement
with Civis Analytics.  This work was supported by the University of
Illinois System Office, the Office of the Vice-Chancellor for Research
and Innovation, the Grainger College of Engineering, and the Department
of Physics at the University of Illinois at Urbana-Champaign. Z.J.W. is
supported in part by the United States Department of Energy
Computational Science Graduate Fellowship, provided under Grant No.
DE-FG02-97ER25308. This work made use of the Illinois Campus Cluster, a
computing resource that is operated by the Illinois Campus Cluster
Program (ICCP) in conjunction with the National Center for
Supercomputing Applications (NCSA) and which is supported by funds from
the University of Illinois at Urbana-Champaign. This research was
partially done at, and used resources of the Center for Functional
Nanomaterials, which is a U.S. DOE Office of Science Facility, at
Brookhaven National Laboratory under Contract No.~DE-SC0012704.
\end{acknowledgments}

\bibliographystyle{IEEEtran}
\bibliography{main}

% Generated by IEEEtran.bst, version: 1.14 (2015/08/26)
\begin{thebibliography}{10}
\providecommand{\url}[1]{#1}
\csname url@samestyle\endcsname
\providecommand{\newblock}{\relax}
\providecommand{\bibinfo}[2]{#2}
\providecommand{\BIBentrySTDinterwordspacing}{\spaceskip=0pt\relax}
\providecommand{\BIBentryALTinterwordstretchfactor}{4}
\providecommand{\BIBentryALTinterwordspacing}{\spaceskip=\fontdimen2\font plus
\BIBentryALTinterwordstretchfactor\fontdimen3\font minus
  \fontdimen4\font\relax}
\providecommand{\BIBforeignlanguage}[2]{{%
\expandafter\ifx\csname l@#1\endcsname\relax
\typeout{** WARNING: IEEEtran.bst: No hyphenation pattern has been}%
\typeout{** loaded for the language `#1'. Using the pattern for}%
\typeout{** the default language instead.}%
\else
\language=\csname l@#1\endcsname
\fi
#2}}
\providecommand{\BIBdecl}{\relax}
\BIBdecl

\bibitem{Illinoisresponsewiki}
All executive orders related to COVID-19 issued by the State of Illinois
  Governor JB Pritzker are listed here:
  \url{https://www2.illinois.gov/government/executive-orders}. The timeline of
  COVID-19 mitigation in Illinois is summarized at:
  \url{https://en.wikipedia.org/wiki/COVID-19_pandemic_in_Illinois\#Government_response}.

\bibitem{keeling2011modeling}
M.~J. Keeling and P.~Rohani, \emph{{Modeling infectious diseases in humans and
  animals}}.\hskip 1em plus 0.5em minus 0.4em\relax Princeton University Press,
  2011.

\bibitem{Murray2020}
C.~J.~L. Murray, ``{Forecasting COVID-19 impact on hospital bed-days, ICU-days,
  ventilator-days and deaths by US state in the next 4 months},''
  \emph{medRxiv}, p. 2020.03.27.20043752, 2020.

\bibitem{MA2020129}
\BIBentryALTinterwordspacing
J.~Ma, ``{Estimating epidemic exponential growth rate and basic reproduction
  number},'' \emph{Infectious Disease Modelling}, vol.~5, pp. 129 -- 141, 2020.
  [Online]. Available:
  \url{http://www.sciencedirect.com/science/article/pii/S2468042719300491}
\BIBentrySTDinterwordspacing

\bibitem{wu2020nowcasting}
J.~T. Wu, K.~Leung, and G.~M. Leung, ``Nowcasting and forecasting the potential
  domestic and international spread of the 2019-ncov outbreak originating in
  wuhan, china: a modelling study,'' \emph{The Lancet}, vol. 395, no. 10225,
  pp. 689--697, 2020.

\bibitem{Fraser2007}
\BIBentryALTinterwordspacing
C.~Fraser, ``{Estimating individual and household reproduction numbers in an
  emerging epidemic},'' \emph{PLoS One}, vol.~2, no.~8, p. e758, 2007.
  [Online]. Available: \url{https://www.ncbi.nlm.nih.gov/pubmed/17712406}
\BIBentrySTDinterwordspacing

\bibitem{chowell2017fitting}
G.~Chowell, ``Fitting dynamic models to epidemic outbreaks with quantified
  uncertainty: a primer for parameter uncertainty, identifiability, and
  forecasts,'' \emph{Infectious Disease Modelling}, vol.~2, no.~3, pp.
  379--398, 2017.

\bibitem{Wu2020}
\BIBentryALTinterwordspacing
J.~T. Wu, K.~Leung, M.~Bushman, N.~Kishore, R.~Niehus, P.~M. de~Salazar, B.~J.
  Cowling, M.~Lipsitch, and G.~M. Leung, ``{Estimating clinical severity of
  COVID-19 from the transmission dynamics in Wuhan, China},'' \emph{Nat Med},
  vol.~26, no.~4, pp. 506--510, 2020. [Online]. Available:
  \url{https://www.ncbi.nlm.nih.gov/pubmed/32284616}
\BIBentrySTDinterwordspacing

\bibitem{viboud2006synchrony}
C.~Viboud, O.~N. Bj{\o}rnstad, D.~L. Smith, L.~Simonsen, M.~A. Miller, and
  B.~T. Grenfell, ``Synchrony, waves, and spatial hierarchies in the spread of
  influenza,'' \emph{Science}, vol. 312, no. 5772, pp. 447--451, 2006.

\bibitem{LloydSmith2005}
\BIBentryALTinterwordspacing
J.~O. Lloyd-Smith, S.~J. Schreiber, P.~E. Kopp, and W.~M. Getz,
  ``{Superspreading and the effect of individual variation on disease
  emergence},'' \emph{Nature}, vol. 438, no. 7066, pp. 355--9, 2005. [Online].
  Available: \url{https://www.ncbi.nlm.nih.gov/pubmed/16292310}
\BIBentrySTDinterwordspacing

\bibitem{small2006super}
M.~Small, C.~Tse, and D.~M. Walker, ``Super-spreaders and the rate of
  transmission of the sars virus,'' \emph{Physica D: Nonlinear Phenomena}, vol.
  215, no.~2, pp. 146--158, 2006.

\bibitem{bansal2007individual}
S.~Bansal, B.~T. Grenfell, and L.~A. Meyers, ``When individual behaviour
  matters: homogeneous and network models in epidemiology,'' \emph{Journal of
  the Royal Society Interface}, vol.~4, no.~16, pp. 879--891, 2007.

\bibitem{kim2018agent}
Y.~Kim, H.~Ryu, and S.~Lee, ``Agent-based modeling for super-spreading events:
  A case study of mers-cov transmission dynamics in the republic of korea,''
  \emph{International Journal of Environmental Research and Public Health},
  vol.~15, no.~11, p. 2369, 2018.

\bibitem{dezsHo2002halting}
Z.~Dezs{\H{o}} and A.-L. Barab{\'a}si, ``Halting viruses in scale-free
  networks,'' \emph{Physical Review E}, vol.~65, no.~5, p. 055103, 2002.

\bibitem{rock2014dynamics}
K.~Rock, S.~Brand, J.~Moir, and M.~J. Keeling, ``Dynamics of infectious
  diseases,'' \emph{Reports on Progress in Physics}, vol.~77, no.~2, p. 026602,
  2014.

\bibitem{pastor2015epidemic}
R.~Pastor-Satorras, C.~Castellano, P.~Van~Mieghem, and A.~Vespignani,
  ``Epidemic processes in complex networks,'' \emph{Reviews of Modern Physics},
  vol.~87, no.~3, p. 925, 2015.

\bibitem{kermack1927contribution}
W.~O. Kermack and A.~G. McKendrick, ``A contribution to the mathematical theory
  of epidemics,'' \emph{Proceedings of the Royal Society of London. Series A,
  Containing papers of a mathematical and physical character}, vol. 115, no.
  772, pp. 700--721, 1927.

\bibitem{Wallinga2006}
J.~Wallinga and M.~Lipsitch, ``{How generation intervals shape the relationship
  between growth rates and reproductive numbers},'' \emph{Proceedings of the
  Royal Society B: Biological Sciences}, vol. 274, no. 1609, pp. 599--604,
  2006.

\bibitem{Neher2020}
R.~A. Neher, R.~Dyrdak, V.~Druelle, E.~B. Hodcroft, and J.~Albert, ``{Potential
  impact of seasonal forcing on a SARS-CoV-2 pandemic},'' \emph{Swiss Medical
  Weekly}, vol. 150, no. 1112, 2020.

\bibitem{smirnova2019forecasting}
A.~Smirnova, L.~deCamp, and G.~Chowell, ``{Forecasting epidemics through
  nonparametric estimation of time-dependent transmission rates using the SEIR
  model},'' \emph{Bulletin of Mathematical Biology}, vol.~81, no.~11, pp.
  4343--4365, 2019.

\bibitem{Ku2020}
C.~C. Ku, T.-C. Ng, and H.-H. Lin, ``{Epidemiological Benchmarks of the
  COVID-19 Outbreak Control in China after Wuhan's Lockdown: A Modelling Study
  with An Empirical Approach},'' \emph{SSRN Electronic Journal}, 2020.

\bibitem{mena2020using}
R.~H. Mena, J.~X. Velasco-Hernandez, N.~B. Mantilla-Beniers, G.~A.
  Carranco-Sapi{\'e}ns, L.~Benet, D.~Boyer, and I.~P. Castillo, ``{Using the
  posterior predictive distribution to analyse epidemic models: COVID-19 in
  Mexico City},'' \emph{arXiv preprint arXiv:2005.02294}, 2020.

\bibitem{gatto2020spread}
M.~Gatto, E.~Bertuzzo, L.~Mari, S.~Miccoli, L.~Carraro, R.~Casagrandi, and
  A.~Rinaldo, ``{Spread and dynamics of the COVID-19 epidemic in Italy: Effects
  of emergency containment measures},'' \emph{Proceedings of the National
  Academy of Sciences}, vol. 117, no.~19, pp. 10\,484--10\,491, 2020.

\bibitem{nishiura_serial}
H.~Nishiura, N.~M. Linton, and A.~R. Akhmetzhanov, ``{Serial interval of novel
  coronavirus (COVID-19) infections},'' \emph{International Journal of
  Infectious Diseases}, vol.~93, pp. 284--286, 2020.

\bibitem{Du2020}
\BIBentryALTinterwordspacing
Z.~Du, X.~Xu, Y.~Wu, L.~Wang, B.~J. Cowling, and L.~A. Meyers, ``{Serial
  Interval of COVID-19 among Publicly Reported Confirmed Cases},'' \emph{Emerg
  Infect Dis}, vol.~26, no.~6, p. 2020.02.19.20025452, 2020. [Online].
  Available: \url{https://www.ncbi.nlm.nih.gov/pubmed/32191173}
\BIBentrySTDinterwordspacing

\bibitem{incub1}
S.~A. Lauer, K.~H. Grantz, Q.~Bi, F.~K. Jones, Q.~Zheng, H.~R. Meredith, A.~S.
  Azman, N.~G. Reich, and J.~Lessler, ``{The Incubation Period of Coronavirus
  Disease 2019 (COVID-19) From Publicly Reported Confirmed Cases: Estimation
  and Application},'' \emph{Annals of Internal Medicine}, vol. 172, no.~9, p.
  577, 2020.

\bibitem{incub2}
N.~M. Linton, T.~Kobayashi, Y.~Yang, K.~Hayashi, A.~R. Akhmetzhanov, S.-m.
  Jung, B.~Yuan, R.~Kinoshita, and H.~Nishiura, ``{Incubation Period and Other
  Epidemiological Characteristics of 2019 Novel Coronavirus Infections with
  Right Truncation: A Statistical Analysis of Publicly Available Case Data},''
  \emph{Journal of Clinical Medicine}, vol.~9, no.~2, p. 538, 2020.

\bibitem{Verity2020}
\BIBentryALTinterwordspacing
R.~Verity, L.~C. Okell, I.~Dorigatti, P.~Winskill, C.~Whittaker, N.~Imai,
  G.~Cuomo-Dannenburg, H.~Thompson, P.~G.~T. Walker, H.~Fu, A.~Dighe, J.~T.
  Griffin, M.~Baguelin, S.~Bhatia, A.~Boonyasiri, A.~Cori, Z.~Cucunuba,
  R.~FitzJohn, K.~Gaythorpe, W.~Green, A.~Hamlet, W.~Hinsley, D.~Laydon,
  G.~Nedjati-Gilani, S.~Riley, S.~van Elsland, E.~Volz, H.~Wang, Y.~Wang,
  X.~Xi, C.~A. Donnelly, A.~C. Ghani, and N.~M. Ferguson, ``{Estimates of the
  severity of coronavirus disease 2019: a model-based analysis},'' \emph{Lancet
  Infect Dis}, 2020. [Online]. Available:
  \url{https://www.ncbi.nlm.nih.gov/pubmed/32240634}
\BIBentrySTDinterwordspacing

\bibitem{Bhatraju2020}
\BIBentryALTinterwordspacing
P.~K. Bhatraju, B.~J. Ghassemieh, M.~Nichols, R.~Kim, K.~R. Jerome, A.~K.
  Nalla, A.~L. Greninger, S.~Pipavath, M.~M. Wurfel, L.~Evans, P.~A. Kritek,
  T.~E. West, A.~Luks, A.~Gerbino, C.~R. Dale, J.~D. Goldman, S.~O'Mahony, and
  C.~Mikacenic, ``{Covid-19 in Critically Ill Patients in the Seattle Region -
  Case Series},'' \emph{N Engl J Med}, 2020. [Online]. Available:
  \url{https://www.ncbi.nlm.nih.gov/pubmed/32227758}
\BIBentrySTDinterwordspacing

\bibitem{Wuhan_clinical}
X.~Yang, Y.~Yu, J.~Xu, H.~Shu, J.~Xia, H.~Liu, Y.~Wu, L.~Zhang, Z.~Yu, M.~Fang,
  T.~Yu, Y.~Wang, S.~Pan, X.~Zou, S.~Yuan, and Y.~Shang, ``{Clinical course and
  outcomes of critically ill patients with SARS-CoV-2 pneumonia in Wuhan,
  China: a single-centered, retrospective, observational study},'' \emph{The
  Lancet Respiratory Medicine}, vol.~8, no.~5, pp. 475--481, 2020.

\bibitem{icnarc}
``{ICNARC report on COVID-19 in critical care, 08 May 2020},'' 2020,
  \url{https://www.icnarc.org/Our-Audit/Audits/Cmp/Reports}.

\bibitem{systematic_review_ifr}
G.~Meyerowitz-Katz and L.~Merone, ``{A systematic review and meta-analysis of
  published research data on COVID-19 infection-fatality rates},''
  \emph{medRxiv}, p. 2020.05.03.20089854, 2020.

\bibitem{emcee}
D.~{Foreman-Mackey}, W.~{Farr}, M.~{Sinha}, A.~{Archibald}, D.~{Hogg},
  J.~{Sanders}, J.~{Zuntz}, P.~{Williams}, A.~{Nelson}, M.~{de Val-Borro},
  T.~{Erhardt}, I.~{Pashchenko}, and O.~{Pla}, ``{emcee v3: A Python ensemble
  sampling toolkit for affine-invariant MCMC},'' \emph{The Journal of Open
  Source Software}, vol.~4, no.~43, p. 1864, Nov. 2019.

\bibitem{terBraak2008}
\BIBentryALTinterwordspacing
C.~J.~F. ter Braak and J.~A. Vrugt, ``{Differential Evolution Markov Chain with
  snooker updater and fewer chains},'' \emph{Statistics and Computing},
  vol.~18, no.~4, pp. 435--446, Dec 2008. [Online]. Available:
  \url{https://doi.org/10.1007/s11222-008-9104-9}
\BIBentrySTDinterwordspacing

\bibitem{Braak2006}
\BIBentryALTinterwordspacing
C.~J.~F. ter Braak, ``{A Markov Chain Monte Carlo version of the genetic
  algorithm Differential Evolution: easy Bayesian computing for real parameter
  spaces},'' \emph{Statistics and Computing}, vol.~16, no.~3, pp. 239--249, Sep
  2006. [Online]. Available: \url{https://doi.org/10.1007/s11222-006-8769-1}
\BIBentrySTDinterwordspacing

\bibitem{farr2014more}
B.~Farr, V.~Kalogera, and E.~Luijten, ``A more efficient approach to
  parallel-tempered markov-chain monte carlo for the highly structured
  posteriors of gravitational-wave signals,'' \emph{Physical Review D},
  vol.~90, no.~2, p. 024014, 2014.

\bibitem{IDPH}
{Data} were downloaded from
  \url{https://www.dph.illinois.gov/covid19/covid19-statistics}.

\bibitem{korolev2020identification}
\BIBentryALTinterwordspacing
I.~Korolev, ``{Identification and Estimation of the SEIRD Epidemic Model for
  COVID-19},'' 2020. [Online]. Available:
  \url{https://ssrn.com/abstract=3569367}
\BIBentrySTDinterwordspacing

\bibitem{castro2020predictability}
M.~Castro, S.~Ares, J.~A. Cuesta, and S.~Manrubia, ``{Predictability: Can the
  turning point and end of an expanding epidemic be precisely forecast?}''
  \emph{arXiv}, pp. arXiv--2004, 2020.

\bibitem{roda2020difficult}
W.~C. Roda, M.~B. Varughese, D.~Han, and M.~Y. Li, ``{Why is it difficult to
  accurately predict the COVID-19 epidemic?}'' \emph{Infectious Disease
  Modelling}, vol.~5, p. 271, 2020.

\bibitem{EMS_map}
{See} map at
  \url{https://www.dph.illinois.gov/sites/default/files/resources/ems-regions-map.pdf}.

\bibitem{weitz2020}
\BIBentryALTinterwordspacing
S.~J. Beckett, M.~Dominguez-Mirazo, S.~Lee, C.~Andris, and J.~S. Weitz,
  ``{Spread of COVID-19 through Georgia, USA. Near-term projections and impacts
  of social distancing via a metapopulation model}.'' [Online]. Available:
  \url{https://github.com/WeitzGroup/MAGEmodel_covid19_GA/blob/master/Report/GA_COVID19_assessment_21Apr2020.pdf}
\BIBentrySTDinterwordspacing

\bibitem{Goog}
\url{https://www.blog.google/technology/health/covid-19-community-mobility-reports?hl=en}.

\bibitem{Unacast}
\url{https://www.unacast.com/post/rounding-out-the-social-distancing-scoreboard}.

\bibitem{flaxman2020report}
\BIBentryALTinterwordspacing
S.~Flaxman, S.~Mishra, A.~Gandy, H.~Unwin, H.~Coupland, T.~Mellan
  \emph{et~al.}, ``{Report 13. Estimating the number of infections and the
  impact of non-pharmaceutical interventions on COVID-19 in 11 European
  countries. Imperial College London, 2020},'' 2020. [Online]. Available:
  \url{https://doi.org/10.25561/77731}
\BIBentrySTDinterwordspacing

\bibitem{Juliette2020report}
H.~J.~T. Unwin, S.~Mishra, V.~Bradley, A.~Gandy, M.~Vollmer, T.~Mellan,
  H.~Coupland, K.~Ainslie, C.~Whittaker, J.~Ish-Horowicz, S.~Filippi, X.~Xi,
  M.~Monod, O.~Ratmann, M.~Hutchinson, F.~Valka, H.~Zhu, I.~Hawryluk,
  P.~Milton, M.~Baguelin, A.~Boonyasiri, N.~Brazeau, L.~Cattarino, G.~Charles,
  L.~V. Cooper, Z.~Cucunuba, G.~CuomoDannenburg, B.~Djaafara, I.~Dorigatti,
  O.~J. Eales, J.~Eaton, S.~van Elsland, R.~FitzJohn, K.~Gaythorpe, W.~Green,
  T.~Hallett, W.~Hinsley, N.~Imai, B.~Jeffrey, E.~Knock, D.~Laydon, J.~Lees,
  G.~Nedjati-Gilani, P.~Nouvellet, L.~Okell, A.~Ower, K.~V. Parag, I.~Siveroni,
  H.~A. Thompson, R.~Verity, P.~Walker, C.~Walters, Y.~Wang, O.~J. Watson,
  L.~Whittles, A.~Ghani, N.~M. Ferguson, S.~Riley, C.~A. Donnelly, S.~Bhat, and
  S.~Flaxman, ``{Report 23: State-level tracking of COVID-19 in the United
  States WHO Collaborating Centre for Infectious Disease Modelling MRC Centre
  for Global Infectious Disease Analytics},'' 2020.

\bibitem{Maslov2020}
\BIBentryALTinterwordspacing
S.~Maslov and N.~Goldenfeld, ``{Window of Opportunity for Mitigation to Prevent
  Overflow of ICU capacity in Chicago by COVID-19},'' \emph{medRxiv}, 2020.
  [Online]. Available:
  \url{https://www.medrxiv.org/content/early/2020/03/24/2020.03.20.20040048}
\BIBentrySTDinterwordspacing

\bibitem{morris2020optimal}
D.~H. Morris, F.~W. Rossine, J.~B. Plotkin, and S.~A. Levin, ``Optimal,
  near-optimal, and robust epidemic control,'' \emph{arXiv preprint
  arXiv:2004.02209}, 2020.

\bibitem{hellewell2020feasibility}
J.~Hellewell, S.~Abbott, A.~Gimma, N.~I. Bosse, C.~I. Jarvis, T.~W. Russell,
  J.~D. Munday, A.~J. Kucharski, and R.~M. Eggo, ``{Feasibility of controlling
  COVID-19 outbreaks by isolation of cases and contacts},'' \emph{The Lancet
  Global Health}, 2020.

\bibitem{Ferretti2020}
\BIBentryALTinterwordspacing
L.~Ferretti, C.~Wymant, M.~Kendall, L.~Zhao, A.~Nurtay, L.~Abeler-Dorner,
  M.~Parker, D.~Bonsall, and C.~Fraser, ``{Quantifying SARS-CoV-2 transmission
  suggests epidemic control with digital contact tracing},'' \emph{Science}, p.
  eabb6936, 2020. [Online]. Available:
  \url{https://www.ncbi.nlm.nih.gov/pubmed/32234805}
\BIBentrySTDinterwordspacing

\bibitem{GWU}
{Fitzhugh Mullan Institute for Health Workforce Equity, George Washington
  University}, ``{Contact tracing workforce estimator},''
  \url{https://www.gwhwi.org/estimator-613404.html}.

\bibitem{Holmdahl2020}
\BIBentryALTinterwordspacing
I.~Holmdahl and C.~Buckee, ``{Wrong but Useful — What Covid-19 Epidemiologic
  Models Can and Cannot Tell Us},'' \emph{New England Journal of Medicine},
  2020. [Online]. Available: \url{https://doi.org/10.1056/NEJMp2016822}
\BIBentrySTDinterwordspacing

\bibitem{stark2006concepts}
K.~D. St{\"a}rk, G.~Regula, J.~Hernandez, L.~Knopf, K.~Fuchs, R.~S. Morris, and
  P.~Davies, ``{Concepts for risk-based surveillance in the field of veterinary
  medicine and veterinary public health: review of current approaches},''
  \emph{BMC health services research}, vol.~6, no.~1, p.~20, 2006.

\bibitem{foddai2020surveillance}
A.~Foddai, A.~Lindberg, J.~Lubroth, and J.~Ellis-Iversen, ``{Surveillance to
  improve evidence for community control decisions during the COVID-19
  pandemic--Opening the animal epidemic toolbox for Public Health},'' \emph{One
  Health}, vol.~9, p. 100130, 2020.

\bibitem{Thecityrepo}
Data were downloaded from \url{https://github.com/thecityny/covid-19-nyc-data}.

\bibitem{Thecityny}
Data originally due to \url{https://www.thecity.nyc/}.

\bibitem{Nychealth}
Data were downloaded from \url{https://github.com/nychealth/coronavirus-data}.

\bibitem{Liu2020}
\BIBentryALTinterwordspacing
Y.~Liu, R.~M. Eggo, and A.~J. Kucharski, ``{Secondary attack rate and
  superspreading events for SARS-CoV-2},'' \emph{Lancet}, vol. 395, no. 10227,
  p. e47, 2020. [Online]. Available:
  \url{https://www.ncbi.nlm.nih.gov/pubmed/32113505}
\BIBentrySTDinterwordspacing

\bibitem{Raines2020.05.23.20111278}
\BIBentryALTinterwordspacing
K.~S. Raines, S.~Doniach, and G.~Bhanot, ``The transmission of sars-cov-2 is
  likely comodulated by temperature and by relative humidity,'' \emph{medRxiv},
  2020. [Online]. Available:
  \url{https://www.medrxiv.org/content/early/2020/05/26/2020.05.23.20111278}
\BIBentrySTDinterwordspacing

\bibitem{Baker2020.04.03.20052787}
\BIBentryALTinterwordspacing
R.~E. Baker, W.~Yang, G.~A. Vecchi, C.~J.~E. Metcalf, and B.~T. Grenfell,
  ``Susceptible supply limits the role of climate in the covid-19 pandemic,''
  \emph{medRxiv}, 2020. [Online]. Available:
  \url{https://www.medrxiv.org/content/early/2020/04/07/2020.04.03.20052787}
\BIBentrySTDinterwordspacing

\bibitem{King2015}
A.~A. King, M.~Domenech~de Cell{\`e}s, F.~M. Magpantay, and P.~Rohani,
  ``{Avoidable errors in the modelling of outbreaks of emerging pathogens, with
  special reference to Ebola},'' \emph{Proceedings of the Royal Society B:
  Biological Sciences}, vol. 282, no. 1806, p. 20150347, 2015.

\bibitem{mckane2005predator}
A.~J. McKane and T.~J. Newman, ``{Predator-prey cycles from resonant
  amplification of demographic stochasticity},'' \emph{Physical Review
  Letters}, vol.~94, no.~21, p. 218102, 2005.

\bibitem{butler2009robust}
T.~Butler and N.~Goldenfeld, ``{Robust ecological pattern formation induced by
  demographic noise},'' \emph{Physical Review E}, vol.~80, no.~3, p. 030902,
  2009.

\bibitem{karig2018stochastic}
D.~Karig, K.~M. Martini, T.~Lu, N.~A. DeLateur, N.~Goldenfeld, and R.~Weiss,
  ``{Stochastic Turing patterns in a synthetic bacterial population},''
  \emph{Proceedings of the National Academy of Sciences}, vol. 115, no.~26, pp.
  6572--6577, 2018.

\bibitem{python}
G.~Van~Rossum and F.~L. Drake, \emph{Python 3 Reference Manual}.\hskip 1em plus
  0.5em minus 0.4em\relax Scotts Valley, CA: CreateSpace, 2009.

\bibitem{numpy}
T.~E. Oliphant, \emph{A guide to NumPy}.\hskip 1em plus 0.5em minus 0.4em\relax
  Trelgol Publishing USA, 2006, vol.~1.

\bibitem{scipy}
P.~{Virtanen}, R.~{Gommers}, T.~E. {Oliphant}, M.~{Haberland}, T.~{Reddy},
  D.~{Cournapeau}, E.~{Burovski}, P.~{Peterson}, W.~{Weckesser}, J.~{Bright},
  S.~J. {van der Walt}, M.~{Brett}, J.~{Wilson}, K.~{Jarrod Millman},
  N.~{Mayorov}, A.~R.~J. {Nelson}, E.~{Jones}, R.~{Kern}, E.~{Larson},
  C.~{Carey}, {\.I}.~{Polat}, Y.~{Feng}, E.~W. {Moore}, J.~{Vand erPlas},
  D.~{Laxalde}, J.~{Perktold}, R.~{Cimrman}, I.~{Henriksen}, E.~A. {Quintero},
  C.~R. {Harris}, A.~M. {Archibald}, A.~H. {Ribeiro}, F.~{Pedregosa}, P.~{van
  Mulbregt}, and S.~.~. {Contributors}, ``{SciPy 1.0: Fundamental Algorithms
  for Scientific Computing in Python},'' \emph{Nature Methods}, 2020.

\bibitem{pandas}
\BIBentryALTinterwordspacing
T.~pandas~development team, ``pandas-dev/pandas: Pandas,'' Feb. 2020. [Online].
  Available: \url{https://doi.org/10.5281/zenodo.3509134}
\BIBentrySTDinterwordspacing

\bibitem{cornerpy}
\BIBentryALTinterwordspacing
D.~Foreman-Mackey, ``corner.py: Scatterplot matrices in python,'' \emph{The
  Journal of Open Source Software}, vol.~24, 2016. [Online]. Available:
  \url{http://dx.doi.org/10.5281/zenodo.45906}
\BIBentrySTDinterwordspacing

\bibitem{matplotlib}
J.~D. Hunter, ``Matplotlib: A 2d graphics environment,'' \emph{Computing in
  Science \& Engineering}, vol.~9, no.~3, pp. 90--95, 2007.

\bibitem{chinacdcweekly}
\BIBentryALTinterwordspacing
{The Novel Coronavirus Pneumonia Emergency Response Epidemiology Team}, ``{The
  Epidemiological Characteristics of an Outbreak of 2019 Novel Coronavirus
  Diseases (COVID-19) — China, 2020},'' \emph{China CDC Weekly}, vol.~2, p.
  113, 2020. [Online]. Available:
  \url{http://weekly.chinacdc.cn//article/id/e53946e2-c6c4-41e9-9a9b-fea8db1a8f51}
\BIBentrySTDinterwordspacing

\bibitem{Mizumoto2020}
\BIBentryALTinterwordspacing
K.~Mizumoto, K.~Kagaya, A.~Zarebski, and G.~Chowell, ``{Estimating the
  asymptomatic proportion of coronavirus disease 2019 (COVID-19) cases on board
  the Diamond Princess cruise ship, Yokohama, Japan, 2020},''
  \emph{Eurosurveillance}, vol.~25, no.~10, 2020. [Online]. Available:
  \url{https://www.eurosurveillance.org/content/10.2807/1560-7917.ES.2020.25.10.2000180}
\BIBentrySTDinterwordspacing

\bibitem{Nishiura2020}
\BIBentryALTinterwordspacing
H.~Nishiura, T.~Kobayashi, T.~Miyama, A.~Suzuki, S.~Jung, K.~Hayashi,
  R.~Kinoshita, Y.~Yang, B.~Yuan, A.~R. Akhmetzhanov, and N.~M. Linton,
  ``{Estimation of the asymptomatic ratio of novel coronavirus infections
  (COVID-19)},'' 2020. [Online]. Available:
  \url{https://www.medrxiv.org/content/early/2020/02/17/2020.02.03.20020248}
\BIBentrySTDinterwordspacing

\bibitem{unagedist}
{Data} were downloaded from \url{https://data.un.org/}.

\end{thebibliography}
\clearpage
\onecolumngrid

\section*{Supplementary Material}

\SupplementaryMaterials

We detail the parameterization of our age-dependent severity model below.
As noted in Section~\ref{subsec:model-time-description}, we track the number of individuals in different age groups separately, and thus we specify the fiducial transition
probabilities for each age group independently. We further define an overall scaling prefactor $p_x$ that governs the final transition probability $p_{x, i}$ as the product of the fiducial rate and the prefactor. Our fiducial transition rates $p_{x,i} / p_x$ are drawn from Ref.~\cite{Verity2020,chinacdcweekly,Mizumoto2020,Nishiura2020}.

In Table~\ref{tab:severity-model} we list the age-dependent values for the probabilities that: infected individuals experience symptoms
$p_{\sigma, i} / p_\sigma$, symptomatic individuals are hospitalized $p_{h, i} / p_h$,
hospitalized patients enter the ICU $p_{c, i} / p_c$, and ICU patients expire $p_{d, i} / p_d$. We also list the relative age distribution of the population in the United States, which we source from the UN~\cite{unagedist}. As specified in Table~\ref{tab:sample-params}, we sample over the $p_c$ and $p_d$ scale factors.

\begin{table*}[ht!]
\caption{
    Table of age-specific parameters, including the age distribution $w_i$
    and the various probabilities of transitioning between various states in our model.
    We present the unscaled distribution $p_{x, i} / p_x$, where $p_x$ denotes the overall scaling which is
    sampled during parameter inference (or whose value is otherwise set as described in the text).
}
\begin{ruledtabular}
    \begin{tabular}{llllllllll}
    {} &   $[0, 10)$ &  $[10, 20)$ &  $[20, 30)$ &  $[30, 40)$ &  $[40, 50)$ & $[50, 60)$ &  $[60, 70)$ &   $[70, 80)$ & $[80, \infty)$ \\
    \midrule
    $w_i$                      &  $0.120004$ &  $0.127891$ &  $0.139256$ &  $0.134948$ &  $0.121898$ &  $0.12725$ &  $0.116278$ &  $0.0727565$ &    $0.0397193$ \\
    $p_{\sigma, i} / p_\sigma$ &     $0.057$ &     $0.054$ &     $0.294$ &     $0.668$ &     $0.614$ &     $0.83$ &      $0.99$ &      $0.995$ &        $0.999$ \\
    $p_{h, i} / p_h$           &     $0.001$ &     $0.003$ &     $0.012$ &     $0.032$ &     $0.049$ &    $0.102$ &     $0.166$ &      $0.243$ &        $0.273$ \\
    $p_{c, i} / p_c$           &      $0.05$ &      $0.05$ &      $0.05$ &      $0.05$ &     $0.063$ &    $0.122$ &     $0.274$ &      $0.432$ &        $0.709$ \\
    $p_{d, i} / p_d$           &       $0.3$ &       $0.3$ &       $0.3$ &       $0.3$ &       $0.3$ &      $0.4$ &       $0.4$ &        $0.5$ &          $0.5$ \\
    \end{tabular}
\end{ruledtabular}
\label{tab:severity-model}
\stepcounter{stable}
\end{table*}

Because the data to which we calibrate does not constrain the symptomatic population, we cannot
observe $p_h$ and so we fix it to one. We force our model to produce a given infection fatality ratio by setting $p_\sigma$ to
\begin{align}
    p_\sigma
    &= \frac{\mathrm{IFR}}{\sum_i w_i (p_{\sigma, i} / p_\sigma) p_{c, i} p_{h, i} p_{d, i} F_\mathrm{tot}},
\end{align}
where the value for $p_{\sigma, i} / p_\sigma$ on the right hand side of the equation is taken first from Table~\ref{tab:severity-model}. This equation is valid if the initial infected population is distributed in proportion to $w_i$.

In Table~\ref{table:parameters1}, we summarize
the posterior probability distribution
for all sample parameters enumerated in Table~\ref{tab:sample-params}.

\begin{table*}[ht!]
\caption{
    Table of inferred model parameters for disease severity and the time dependence of mitigation effects.
    Mitigation was modeled with a piecewise Hermite cubic interpolating polynomial
    as described in Section~\ref{subsec:model-time-description}.
    Each column reports median parameter values from the MCMC algorithm for the different model fits
    along with the largest of the upper and lower one-sigma error bounds reported by the algorithm.
    In addition, we report the (derived) values and uncertainties of $R_t$ (as defined in Eq.~\ref{eqn:Rt})
    for each model as evaluated on May 1, 2020.
    The column headers specify the modeled region of Illinois and the inclusive end date of the calibration data.
}
\begin{ruledtabular}
\begin{tabular}{lllllll}
{} &                Illinois (5/17) &                Illinois (4/20) &            Northeastern (5/17) &           North-Central (5/17) &                 Central (5/17) &                Southern (5/17) \\
\midrule
$R_0$                     &                 $2.2 \pm 0.26$ &                 $2.2 \pm 0.25$ &                 $2.3 \pm 0.28$ &                 $1.6 \pm 0.38$ &                  $1.6 \pm 0.4$ &                 $1.6 \pm 0.33$ \\
$t_s$                     &  $\mathrm{Feb} \, 15 \pm 3.92$ &  $\mathrm{Feb} \, 14 \pm 4.24$ &  $\mathrm{Feb} \, 16 \pm 3.73$ &  $\mathrm{Feb} \, 16 \pm 7.82$ &  $\mathrm{Feb} \, 18 \pm 8.25$ &  $\mathrm{Feb} \, 17 \pm 8.40$ \\
$t_0$                     &   $\mathrm{Mar} \, 9 \pm 4.13$ &  $\mathrm{Mar} \, 10 \pm 4.13$ &   $\mathrm{Mar} \, 9 \pm 4.04$ &   $\mathrm{Mar} \, 8 \pm 5.71$ &   $\mathrm{Mar} \, 8 \pm 6.08$ &  $\mathrm{Mar} \, 11 \pm 7.62$ \\
$t_1$                     &  $\mathrm{Mar} \, 20 \pm 2.94$ &  $\mathrm{Mar} \, 21 \pm 3.28$ &  $\mathrm{Mar} \, 20 \pm 3.01$ &  $\mathrm{Mar} \, 24 \pm 6.04$ &  $\mathrm{Mar} \, 25 \pm 6.58$ &   $\mathrm{Apr} \, 4 \pm 5.53$ \\
$M(t_1)$                  &               $0.48 \pm 0.051$ &               $0.47 \pm 0.048$ &               $0.45 \pm 0.048$ &                $0.64 \pm 0.12$ &                $0.61 \pm 0.12$ &               $0.59 \pm 0.099$ \\
$R_t(\mathrm{May\, 1})$   &                $0.94 \pm 0.01$ &               $0.97 \pm 0.037$ &               $0.94 \pm 0.011$ &                $0.99 \pm 0.02$ &               $0.95 \pm 0.023$ &                $0.9 \pm 0.027$ \\
IFR                       &             $0.007 \pm 0.0012$ &            $0.0071 \pm 0.0012$ &             $0.007 \pm 0.0012$ &             $0.007 \pm 0.0012$ &             $0.007 \pm 0.0012$ &             $0.007 \pm 0.0012$ \\
$\tau_h$                  &                    $6.3 \pm 2$ &                    $6.2 \pm 2$ &                    $6.2 \pm 2$ &                  $6.1 \pm 1.8$ &                  $6.2 \pm 1.8$ &                  $6.9 \pm 1.8$ \\
$\sigma_h$                &                  $2.3 \pm 1.4$ &                  $2.2 \pm 1.6$ &                  $2.4 \pm 1.7$ &                  $4.2 \pm 1.9$ &                  $4.2 \pm 1.9$ &                  $4.6 \pm 1.7$ \\
$\tau_{\mathrm{disch}}$   &                    $7 \pm 1.9$ &                  $6.9 \pm 1.9$ &                  $7.1 \pm 1.9$ &                  $6.2 \pm 1.7$ &                  $6.6 \pm 1.8$ &                  $6.8 \pm 1.9$ \\
$\sigma_{\mathrm{disch}}$ &                  $6.1 \pm 1.8$ &                    $5.7 \pm 2$ &                  $6.4 \pm 1.7$ &                  $3.5 \pm 1.7$ &                  $4.6 \pm 1.9$ &                    $5 \pm 1.8$ \\
$p_c$                     &                $0.61 \pm 0.16$ &                $0.67 \pm 0.18$ &                $0.62 \pm 0.17$ &                $0.47 \pm 0.14$ &               $0.32 \pm 0.097$ &                $0.51 \pm 0.14$ \\
$\tau_c$                  &                 $1.4 \pm 0.97$ &                  $1.7 \pm 1.3$ &                 $1.3 \pm 0.93$ &                    $3 \pm 1.6$ &                  $2.2 \pm 1.5$ &                    $2 \pm 1.3$ \\
$\sigma_c$                &                  $2.2 \pm 1.8$ &                    $2 \pm 1.8$ &                  $2.3 \pm 1.9$ &                  $2.5 \pm 1.6$ &                  $2.5 \pm 1.7$ &                  $2.7 \pm 1.7$ \\
$p_d$                     &                 $1.4 \pm 0.15$ &                 $1.3 \pm 0.16$ &                 $1.4 \pm 0.16$ &                  $1.6 \pm 0.3$ &                 $1.9 \pm 0.25$ &                 $1.9 \pm 0.26$ \\
$\tau_{d}$                &                 $8.8 \pm 0.96$ &                    $7.7 \pm 1$ &                 $8.8 \pm 0.98$ &                   $13 \pm 2.1$ &                  $8.9 \pm 1.6$ &                   $11 \pm 1.4$ \\
$\sigma_{d}$              &                  $4.6 \pm 1.5$ &                  $3.7 \pm 1.5$ &                  $4.7 \pm 1.6$ &                  $9.2 \pm 2.4$ &                    $7 \pm 2.7$ &                  $6.6 \pm 2.2$ \\
$\tau_\mathrm{rec}$       &                   $11 \pm 2.3$ &                   $12 \pm 2.5$ &                   $11 \pm 2.3$ &                   $12 \pm 2.8$ &                   $12 \pm 2.8$ &                   $12 \pm 2.8$ \\
$\sigma_\mathrm{rec}$     &                  $6.8 \pm 2.7$ &                  $8.7 \pm 2.9$ &                  $6.9 \pm 2.7$ &                   $10 \pm 2.8$ &                   $10 \pm 2.8$ &                  $9.9 \pm 2.9$ \\
$F_\mathrm{tot}$          &                $1.4 \pm 0.033$ &                 $1.3 \pm 0.06$ &                $1.4 \pm 0.034$ &                 $1.2 \pm 0.14$ &                  $1.6 \pm 0.3$ &                 $2.1 \pm 0.28$ \\
$\tau_\mathrm{tot}$       &                 $2.9 \pm 0.46$ &                   $2 \pm 0.53$ &                 $2.8 \pm 0.47$ &                 $2.2 \pm 0.88$ &                 $2.3 \pm 0.91$ &                 $2.8 \pm 0.88$ \\
$\sigma_{\mathrm{tot}}$   &                 $2.3 \pm 0.88$ &                 $1.9 \pm 0.84$ &                 $2.2 \pm 0.86$ &                 $2.1 \pm 0.91$ &                   $2 \pm 0.92$ &                 $1.9 \pm 0.91$ \\
$A_\mathrm{SF}$           &                $0.15 \pm 0.03$ &               $0.06 \pm 0.071$ &               $0.11 \pm 0.035$ &               $0.13 \pm 0.065$ &                $0.12 \pm 0.07$ &               $0.13 \pm 0.073$ \\
\end{tabular}
\end{ruledtabular}
\label{table:parameters1}
\stepcounter{stable}
\end{table*}

\begin{figure}[!ht]
\includegraphics[width=\columnwidth]{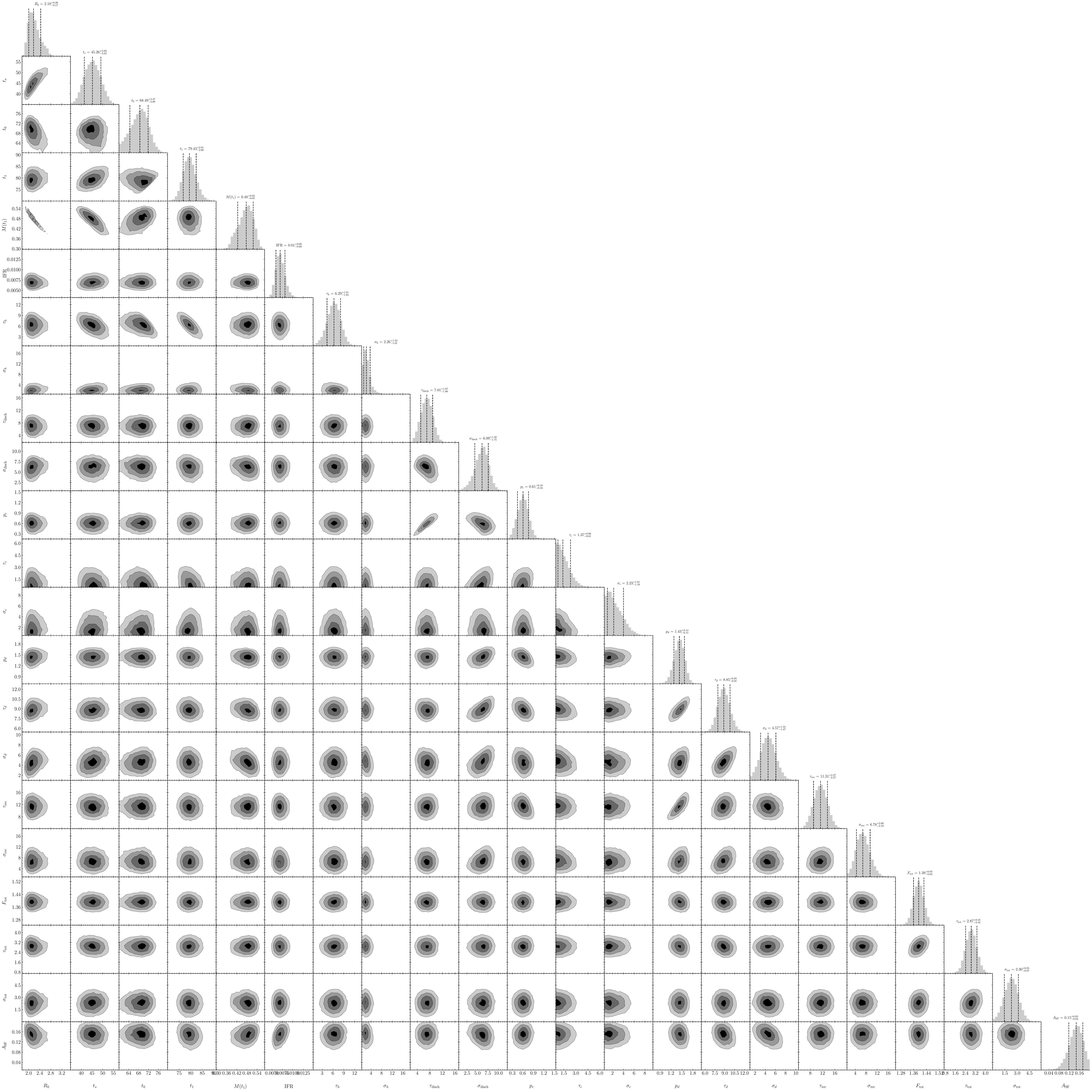}
\caption{Joint posterior distributions of pairs of the complete set of
    parameters of our model fitted to the all-state data up to May 17, 2020.
    The correlations between some pairs of
    fitted parameters such as, e.g., between $R_0$ and that start date of
    the epidemic are reflected in the ellipsoidal shape of the
    posteriors. This is an expanded version of Fig.~\ref{fig:corner}}
\label{fig:fig_S1}
\end{figure}

Figure~\ref{fig:fig_S1} shows joint posterior distributions for all pairs over the complete set of
parameters of our model fitted to all-state data up to May 17, 2020 as shown in the bottom panel of Fig.~\ref{fig2}.

\begin{table}[ht!]
\caption{
    Same as Table~\ref{table:parameters1}, but for fits for New York City.
}
% \begin{ruledtabular}
\begin{tabular}{ll}
\hline
\hline
{} &                     NYC (5/17) \\
\midrule
$R_0$                     &                 $3.4 \pm 0.26$ \\
$t_s$                     &  $\mathrm{Feb} \, 20 \pm 1.69$ \\
$t_0$                     &   $\mathrm{Mar} \, 3 \pm 2.05$ \\
$t_1$                     &  $\mathrm{Mar} \, 24 \pm 2.38$ \\
$M(t_1)$                  &               $0.27 \pm 0.039$ \\
$R_t(\mathrm{May\, 1})$   &               $0.56 \pm 0.042$ \\
IFR                       &           $0.0056 \pm 0.00099$ \\
$\tau_h$                  &                   $10 \pm 1.5$ \\
$\sigma_h$                &                    $6.2 \pm 1$ \\
$\tau_{\mathrm{disch}}$   &                    $2.3 \pm 1$ \\
$\sigma_{\mathrm{disch}}$ &                  $9.4 \pm 1.6$ \\
$p_c$                     &                $0.23 \pm 0.11$ \\
$\tau_c$                  &                 $1.2 \pm 0.31$ \\
$\sigma_c$                &                 $1.9 \pm 0.76$ \\
$p_d$                     &                  $2.1 \pm 0.2$ \\
$\tau_{d}$                &                 $8.3 \pm 0.77$ \\
$\sigma_{d}$              &                     $17 \pm 1$ \\
$\tau_\mathrm{rec}$       &                   $10 \pm 3.5$ \\
$\sigma_\mathrm{rec}$     &                   $12 \pm 3.3$ \\
$A_\mathrm{SF}$           &               $0.13 \pm 0.053$ \\
\hline\hline
\end{tabular}
% \end{ruledtabular}
\label{table:nyc-parameters}
\end{table}

\end{document}